\documentclass[aps,pre,preprint,superscriptaddress]{revtex4}
\usepackage[utf8x]{inputenc}
\usepackage{bm,comment,float}
\usepackage[labelformat=simple]{subcaption}

\usepackage{amsfonts,amssymb,amsmath}
\usepackage{epsfig,doi,cleveref}
\usepackage{graphicx,url,physics}
\usepackage{soul,xcolor,hyperref}

\graphicspath{{./figures/}}

\DeclareMathOperator{\im}{i}

\providecommand{\norm}[1]{\lVert#1\rVert}
\begin{document}
\setstcolor{red}
\title{Phase behavior of a lattice-gas model for biaxial nematics}

\author{W. G. C. Oropesa}
\email[]{carreras@if.usp.br}
\affiliation{Universidade de Sao Paulo, Instituto de Fisica, Rua do Matao, 1371,
05508-090, Sao Paulo, SP, Brazil}
\author{E. S. Nascimento}
\email[]{edusantos18@esp.puc-rio.br}
\affiliation{Dept. of Physics, PUC-Rio, Rua Marquês de São Vicente 225, 22453-900 Rio de Janeiro, Rio de Janeiro, Brazil}
\author{A. P. Vieira}
\email[]{apvieira@if.usp.br}
\affiliation{Universidade de Sao Paulo, Instituto de Fisica, Rua do Matao, 1371,
05508-090, Sao Paulo, SP, Brazil}
\date{\today}
\begin{abstract}

We employ a lattice-gas extension of the Maier--Saupe model with discrete orientation states to study the phase behavior of a statistical model for biaxial nematogenic units in mean-field theory. The phase behavior of the system is investigated in terms of the strength of isotropic interaction between anisotropic objects, as well as the degree of biaxiality and the concentration of those units. We obtain phase diagrams with isotropic phases and stable biaxial and uniaxial nematic structures, various phase coexistences, many types of critical and multicritical behaviors, such as ordinary vapor-liquid critical points, critical end points and tricritical points, and distinct Landau-like multicritical points. Our results widen the possibilities of relating the phenomenological coefficients of the Landau--de Gennes expansion to microscopic parameters, allowing an improved interpretation of theoretical fittings to experimental data.
\end{abstract}

\maketitle

\section{Introduction}\label{sec1}

Nematic mesophases are probably the simplest states of matter observed in liquid-crystalline systems that exhibit long-range orientational order in the absence of translational symmetry breaking \cite{deGennes_Book, FigueiredoNeto2005, Singh2000, Palffy2007}. Indeed, uniaxial nematic structures are characterized macroscopically by the existence of orientation-dependent physical properties (for example, optical or magnetic anisotropies), which lead to the definition of the director of a nematic phase. 
Notwithstanding, the breaking of isotropy in the plane perpendicular to the uniaxial director may lead to the elusive biaxial state, whose possibility was theoretically pointed out by Freiser \cite{freiser1970ordered} about 50 years ago. 
Experimentally, the existence of the biaxial phase was initially 
confirmed for lyotropic systems \cite{yu1980observation}. More recently, there have been claims
of the identification of the phase in thermotropic systems composed of bent-core molecules, although this remains debatable (see Ref.\,\cite{Jakli2018} and references therein). In any case, these claims catalyzed various experimental, computational, and theoretical investigations \cite{Luckhurst2015biaxial,Akpinar2019} of candidate biaxial systems.

Most theoretical and computational studies looking for biaxial phases focus on the orientational order, leaving aside effects associated with a varying density of nematogens. Approaches based on the phenomenological Landau--de Gennes expansion \cite{GRAMSBERGEN1986} are able to partially remedy this situation, by exploiting variations in the expansion coefficients, although these are difficult to connect with microscopic parameters. Our aim in this paper is to investigate the equilibrium phase diagrams of a statistical model in which nematogens with non-cylindrical symmetry can move from site to site in a lattice whose occupation can be controlled. In our model, pairs of nematogens interact via an isotropic potential which can be repulsive or attractive, as well as via an anisotropic potential which favors a biaxial arrangement, leading, at sufficiently high occupation and sufficiently low temperature, to a biaxial phase.

Lattice models of nematic order have a long history in the literature. For uniaxial systems, the pioneering work of Lebwohl and Lasher \cite{Lebwohl1972} inspired a number of other investigations, including a lattice-gas extension by Bates \cite{bates2001computer,bates2002phase}. For biaxial systems, the Luckhurst--Romano model \cite{Luckhurst1980}, based on the truncation of an anisotropic potential to second-rank terms, has been  likewise influential. As a rule, Monte Carlo calculations for nearest-neighbor versions of these models on fully-occupied cubic lattices lead to the same qualitative predictions as those obtained from mean-field versions \cite{maier1958einfache,Longa2005}, despite sometimes substantial quantitative discrepancies \cite{Lebwohl1972,Biscarini1995}. 

A quite general bilinear anisotropic interaction potential $V_{12}$ between two nematogens labeled as $1$ and $2$ was proposed by Straley \cite{Straley1974}. In the two-tensor formulation of Sonnet, Virga and Durand \cite{sonnet2003dielectric}, it takes the form
\begin{equation}
V_{12} = -\frac{9}{4}A\left\{ \mathbf{q}_1\mathbf{:}\mathbf{q}_2 + \zeta\left( \mathbf{q}_1\mathbf{:}\mathbf{b}_2 + \mathbf{b}_1\mathbf{:}\mathbf{q}_2 \right) + \lambda \mathbf{b}_1\mathbf{:}\mathbf{b}_2\right\}.
\label{eq:V12}
\end{equation}
In Eq.\,\eqref{eq:V12}, $A>0$ sets the energy scale, while the second-rank tensors $\mathbf{q}$ and $\mathbf{b}$ are defined in terms of mutually orthogonal unit vectors 
$\hat{n}_1$, $\hat{n}_2$ and $\hat{n}_3$ pointing along the first, second and third principal axes of each nematogen as
\begin{equation}
\mathbf{q}=\hat{n}_{1}\otimes\hat{n}_{1}-\frac{1}{3}\mathbf{I}\quad\text{and}\quad\mathbf{b}=\hat{n}_{2}\otimes\hat{n}_{2}-\hat{n}_{3}\otimes\hat{n}_{3},
\end{equation}
$\mathbf{I}$ being the $3\times3$ identity matrix. 
The operation $\mathbf{q}_1\mathbf{:}\mathbf{q}_2$
is the Frobenius inner product \cite{Horn_Book}, given by $\Tr\left(\mathbf{q}_1\mathbf{q}_2\right)$, where $\Tr\textbf{M}$ is the trace of matrix $\textbf{M}$.
The adimensional parameters $\zeta$ and $\lambda$ gauge the importance of biaxial couplings. If $\zeta=\lambda=0$, Eq.\,\eqref{eq:V12} is reduced to the Maier--Saupe interaction energy \cite{maier1958einfache}, defined solely by the relative orientation of the first principal axes of both nematogens. This is appropriate when dealing with nematogens whose form may be properly approximated as uniaxial. Otherwise, if the nematogens are intrinsically biaxial, a proper description of the interaction energy requires setting either $\zeta$ or $\lambda$ to nonzero values, so that the relative orientations of other principal axes are also relevant. 
Here 
we work with the condition $\lambda=\zeta^2$, corresponding to the London approximation for dispersion forces \cite{sonnet2003dielectric}, which allows us to write $V_{12}$ in the form
\begin{equation}
V_{12} = -\frac{9}{4}A \left(\mathbf{q}_1 + \frac{\Delta}{3}\mathbf{b}_1 \right)\mathbf{:} \left(\mathbf{q}_2 + \frac{\Delta}{3}\mathbf{b}_2 \right).
\label{eq:V12Delta}
\end{equation}
By resorting to a simplified view of a biaxial nematogen as a rectangular platelet, the biaxiality parameter $\Delta = 3\zeta$ can be interpreted in terms of the sides of the platelet, so that $\Delta=0$ would correspond to a ``rod-like'' object, $\Delta=3$ to a ``disk-like'' object and $\Delta=1$ to a maximally biaxial object \cite{nascimento2015maier}.

In the same spirit as the lattice-gas version of the Lebwohl--Lasher model investigated by Bates \cite{bates2001computer}, we allow each site of regular lattice to be empty or occupied by a single nematogen, adding an isotropic interaction to the potential in Eq.\,\eqref{eq:V12Delta} to obtain the contribution of two neighboring sites $i$ and $j$ to the total interaction energy of the system, 
\begin{equation}
V_{ij} = \gamma_i\gamma_j\left\{ U -\frac{9}{4}A \left(\mathbf{q}_i + \frac{\Delta}{3}\mathbf{b}_i \right)\mathbf{:} \left(\mathbf{q}_j + \frac{\Delta}{3}\mathbf{b}_j \right) \right\}.
\label{eq:V12Full}
\end{equation}
The occupation variable $\gamma_i$ is equal to $0$ if site $i$ is empty and to $1$ if the site is occupied. In this work we allow the isotropic interaction parameter $U$ to be either negative, representing attractive interactions, or positive, representing repulsion. This last case could lead to long-range sublattice ordering in cubic lattices, an unphysical feature for a fluid phase. At the mean-field level, however, describing such kind of arrangement would require the explicit introduction of sublattices. Instead, we proceed with the simplest mean-field strategy, which would be appropriate for describing a frustrated lattice or, for that matter, a fluid phase.

In order to perform detailed calculations, besides using Eq.\,\eqref{eq:V12Full} to describe the pair interactions, we also employ the Zwanzig approximation \cite{Zwanzig1963}, which restricts the possible orientations of a nematogen to the coordinate axes. This approximation has been applied in different contexts \cite{de1986reentrant, do2010statistical, do2011phase, liarte2012enhancement, nascimento2015maier, Sauerwein2016, nascimento2016lattice, petri2018field, rodrigues2020magnetic, dosSantos2021}, always leading to qualitative results which fully agree with continuous versions of the corresponding models when a comparison is possible. In particular, when dealing with intrinsically biaxial nematogens, these models are capable of reproducing the qualitative characteristics of nematic phase diagrams, such as sequences of biaxial-uniaxial-isotropic phase transitions with increasing temperature, and a well-defined Landau multicritical point, which signals a direct transition between the isotropic and the biaxial phases \cite{nascimento2016lattice,dosSantos2021}.

Therefore, in this work we investigate the phase diagrams of what may be characterized as a lattice-gas (LG) extension of the Maier--Saupe--Zwanzig model (MSZ), which from now on we will call the LGMSZ model. The LG extension introduces dilution as an extra ingredient in our model, allowing the study of phenomena such as vapor-liquid, vapor-nematic and nematic-nematic (low-high concentration) coexistence. The study of such coexistence is not possible if we treat a model based on a fully-occupied lattice.  

This paper is organized as follows. Sec.\,\ref{sec2} presents the model description and sketches its mean-field solution. In Sec.\,\ref{sec3} we present a detailed analysis of the dilution effects, in the absence of isotropic interactions. Sec.\,\ref{sec4} is dedicated to the study of the effects of isotropic interactions for molecular systems with fixed degrees of biaxiality. In Sec.\,\ref{sec5} we present an analysis of the effects of the biaxiality degree in the multicritical points present in the phase diagrams. Conclusions are drawn in Sec.\,\ref{sec6}. A few technical details are relegated to Appendices \ref{appendix1} and \ref{appendix2}.

\section{The LGMSZ model}\label{sec2}

We consider a lattice system with $N$ sites and  $N_m$ non-spherical objects such that $N \ge N_m$. Each lattice site can be either empty or occupied by an asymmetric object, the state of site $i$ being described by an occupation variable $\gamma_i$ taking the values $0$ (empty site) or $1$ (occupied site). Then, based on Eq.\,\eqref{eq:V12Full}, we define the LGMSZ model by means of effective the Hamiltonian
\begin{equation}\label{eq:MSZUmodel}
    \mathcal{H}=\sum_{(i,j)}V_{ij}=-A\sum_{(i,j)}\gamma_{i}\gamma_{j}\mathbf{\Omega}_{i}\mathbf{:}\mathbf{\Omega}_{j}+U\sum_{(i,j)}\gamma_{i}\gamma_{j},
\end{equation}
where $A$ and $U$ are coupling parameters, with $A>0$, the sum is performed over pairs $(i,j)$ of neighboring sites $i$ and $j$ in the lattice, and the quantity $\mathbf{\Omega}_{i}$ is a second-rank tensor associated with the nematogen at site $i$. Specifically, $\mathbf{\Omega}_{i}$ is represented by a $3\times 3$ square matrix with real entries. For nematogens, $\mathbf{\Omega}_{i}$ is a symmetric traceless matrix, its eigenvalues $\omega_i$ are real and their sum is zero \cite{deGennes_Book,Luckhurst2015biaxial}. Then, we can assume that $\omega_{1}=(-1+\Delta)/2$, $\omega_{2}=(-1-\Delta)/2$, and $\omega_{3}=1$, where the parameter $\Delta$ gauges the asymmetry or biaxiality degree of the object \cite{nascimento2015maier}: $\Delta=0$ for rod-like shapes,  $\Delta=3$ for plate-like shapes, and  $\Delta \neq 0,3$ for brick-like shapes. Biaxial objects with $\Delta=1$ present a maximal degree of asymmetry. Instead of working with continuum orientational states, we follow the Zwanzig prescription \cite{Zwanzig1963} in assuming that the principal axes of a nematogen are restricted to align in the directions of the Cartesian axes, which leads to an effective spin-like model with six states described by diagonal matrices $\bm{\Omega}_i$ \cite{nascimento2015maier}. Notice that, in the limit $\Delta=0$, Eq.\,\eqref{eq:MSZUmodel} reduces to a discretized version of the Lebwohl--Lasher lattice-gas
 model introduced by Bates in Ref.\,\cite{bates2001computer} (with a rescaling of energy, as our parameter $A$ would be equivalent to $2\epsilon/3$, $\epsilon$ being the energy scale of the anisotropic interaction in Ref.\,\cite{bates2001computer}).

The first term in Eq.\,\eqref{eq:MSZUmodel} represents a dilute version of the MSZ model, and the orientation-dependent interaction may give rise to distinct nematic phases. The second term is the isotropic contribution to the pair potential. For the particular case of $U<0$, representing attractive isotropic interactions, one can find phase transitions between isotropic fluid states, in analogy with previous studies \cite{bates2001computer,bates2002phase}. In the present work we assume that the parameter $U$ can also be positive, representing repulsive interactions. In this latter case, as we are interested in modeling fluid phases only, we refrain from trying to account for any kind of sublattice ordering whatsoever. 

Determining the thermodynamic properties of the lattice system defined by Eq. \eqref{eq:MSZUmodel} is rather intricate, due to the complex interplay between the various interactions. Therefore, we think it is appropriate to study the model in a mean-field treatment, which is equivalent to considering the fully-connected Hamiltonian
\begin{equation}\label{MFmodel}
    \mathcal{H}_\text{mf}=-\frac{A}{2N}\sum_{i,j=1}^{N}\gamma_{i}\gamma_{j}\mathbf{\Omega}_{i}\mathbf{:}\mathbf{\Omega}_{j}+ \frac{U}{2N}\sum_{i,j=1}^{N}\gamma_{j}\gamma_{j},
\end{equation}
where the sums over pairs of neighboring sites are replaced by sums over all pairs of sites, and the coupling parameters are replaced by new ones that are inversely proportional to the number of sites to ensure that energy is extensive. 
This form of effective, long-range model has been proposed to investigate the phase behavior of statistical models with nematic-like phases \cite{do2010statistical,nascimento2015maier,nascimento2016lattice,SalinasNascimento2017,
petri2018field,rodrigues2020magnetic}. 
Therefore, our main interest is to study the thermodynamics of phases transitions of the mean-field model in Eq. \eqref{MFmodel}.

The canonical ensemble is the usual route to investigate the macroscopic behavior of Hamiltonian systems in statistical mechanics. Nevertheless, because of its lattice-gas character, the configurations of microscopic variables of our model are subject to the restriction that the sum of $\gamma_i$ over all lattice sites should be equal to $N_m$,
%
%
which leads to complications in evaluating the canonical partition function. As a result, it is more convenient to consider the formalism of the grand canonical ensemble, where the number of nematogens may fluctuate due to the coupling to a particle reservoir \cite{do2010statistical,nascimento2015maier,rodrigues2020magnetic}. Then, we must determine the grand partition function
\begin{equation}\label{eq:gpflgmsz}
    \begin{split}
        \Xi=&\sum_{\{\gamma_{i}\}}\sum_{\{\mathbf{\Omega}_{i}\}}\exp\left(\beta \mathcal{H}_\text{mf} + \beta\mu\sum_{i}\gamma_{i}\right),
    \end{split}
\end{equation}
where $\beta = 1/k_B T$, $k_B$ is the Boltzmann constant (which we take to be equal to $1$ in suitable units), $T$ is the temperature and $\mu$ is the chemical potential.  
In this ensemble, the sum over configurations in Eq. \eqref{eq:gpflgmsz} is no longer restricted, and  mean-field calculations are feasible, as indicated in Appendix \ref{appendix1}.

As a result, we obtain the Landau--de Gennes free-energy functional 
%
\begin{equation}\label{eq:landaufef}
\begin{split}
\psi(S,\eta,\phi) &= \dfrac{A}{4}\left(3S^{2}+\eta^{2}\right)+\dfrac{U}{2}\phi^{2}-\mu\phi\\
 &\quad +\dfrac{1}{\beta}\left[(1-\phi)\ln{\left(\dfrac{1-\phi}{6}\right)}+\phi\ln{(\phi)}\right] \\
 & \quad -\dfrac{\phi}{\beta}\ln{\left[\Lambda(S,\eta)\right]},
\end{split}
\end{equation}
where 
%
\begin{equation}
\begin{split}
\Lambda(S,\eta) &= 2\exp\left[ -\frac{3\beta A}{4}(S+\eta) \right] \cosh{\left[\dfrac{3\beta A}{4}\left(S-\dfrac{\eta}{3}\right)\Delta\right]} \\
&\quad + 2\exp\left[ -\frac{3\beta A}{4}(S-\eta)\right] \cosh{\left[\dfrac{3\beta A}{4}\left(S+\dfrac{\eta}{3}\right)\Delta\right]} \\
	&\quad + 2\exp\left( \frac{3\beta A}{2}S\right) \cosh{\left(\dfrac{\beta A}{2}\eta \Delta\right)}.
\end{split}
\end{equation}
%
and the scalar parameters $S$ and $\eta$ are associated with the symmetric and traceless tensor order parameter \cite{GRAMSBERGEN1986}
\begin{equation} \label{Qtensor}
    \mathbf{Q} = \left< \bm{\Omega} \right> = \frac{1}{2}
    \begin{pmatrix}
    -S - \eta & 0 & 0 \\
    0 & -S + \eta & 0 \\
    0 & 0 &  2S
    \end{pmatrix},
\end{equation}
in which $\left< \cdot \right>$ denotes the ensemble average. 

The equilibrium values of $S$, $\eta$ and $\phi$ are determined by locating the absolute minima of $\psi(S,\eta,\phi)$, leading to the mean-field (MF) equations 
\begin{equation}
\frac{\partial\psi}{\partial S} = \frac{\partial\psi}{\partial\eta} = \frac{\partial\psi}{\partial\phi}=0,
\label{eq:mf}
\end{equation}
which take the self-consistent forms $S=F_{1}(S,\eta,\phi;\beta,\mu,\Delta)$, $\eta=F_{2}(S,\eta,\phi;\beta,\mu,\Delta)$, and $\phi=F_{3}(S,\eta,\phi;\beta,\mu,\Delta)$. 
Depending on the solutions to these mean-field equations, the structure of the eigenvalues ${Q_x,Q_y,Q_z}$ of  $\mathbf{Q}$ may be such that (i) $Q_x=Q_y=Q_z=0$, corresponding to the isotropic phase; (ii) $Q_x=Q_y\neq Q_z$ (or similar relations with permutations of the indices $x$, $y$ and $z$), corresponding to an uniaxial nematic phase; (iii) $Q_x$, $Q_y$ and $Q_z$ all distinct, corresponding to a biaxial nematic phase. If the eigenvalue with the largest absolute value is positive (negative), the nematic solution is calamitic (discotic). We use this terminology for both uniaxial and biaxial cases throughout the paper. In terms of the quantities $S$ and $\eta$, the isotropic solution is given by $S=\eta=0$, uniaxial solutions are such that $S\neq0$ with $\eta=0$ or $\eta = \pm 3S$, while the remaining cases represent biaxial solutions.

We emphasize that the values of $S$, $\eta$ and $\phi$ at the absolute minima of $\psi$ represent thermodynamic equilibrium values for fixed reciprocal temperature $\beta$, chemical potential $\mu$ and biaxiality degree $\Delta$. The free-energy $\mathcal{F}=\mathcal{F}(\beta,\mu,\Delta)$ of the system corresponds to the convex envelope of $\psi$ determined after inserting values of $S$, $\eta$ and $\phi$ associated with the minima of the free-energy functional.


\section{Behavior in the absence of the isotropic interaction}\label{sec3}


We start the investigation by assuming zero isotropic interaction, $U=0$, which simplifies the analysis of the problem by reducing the number of parameters. Some aspects of this case were discussed by Rodrigues \textit{et al.} \cite{rodrigues2020magnetic}, but taking into account only a specific range of model parameters. Here, we will present phase diagrams with many distinct topologies  by exploring a wider range of values of thermodynamic fields. The results with zero isotropic interaction are helpful in understanding the situation involving both isotropic and anisotropic couplings, to be analyzed in the next section.

\begin{figure}[ht]
    \begin{subfigure}[b]{0.45\textwidth}
      \includegraphics[width=\textwidth]{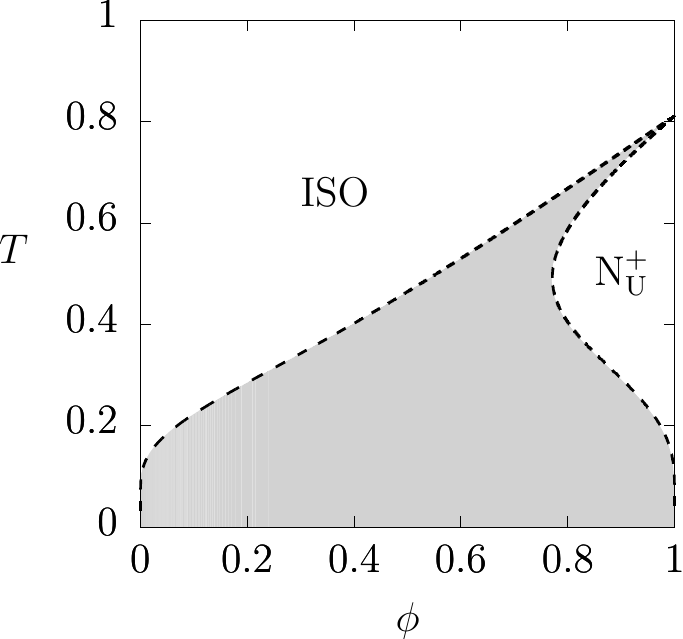}
      \caption{$\Delta=0$}
      \label{fig:1}
    \end{subfigure}\hfill
    \begin{subfigure}[b]{0.45\textwidth}
      \includegraphics[width=\textwidth]{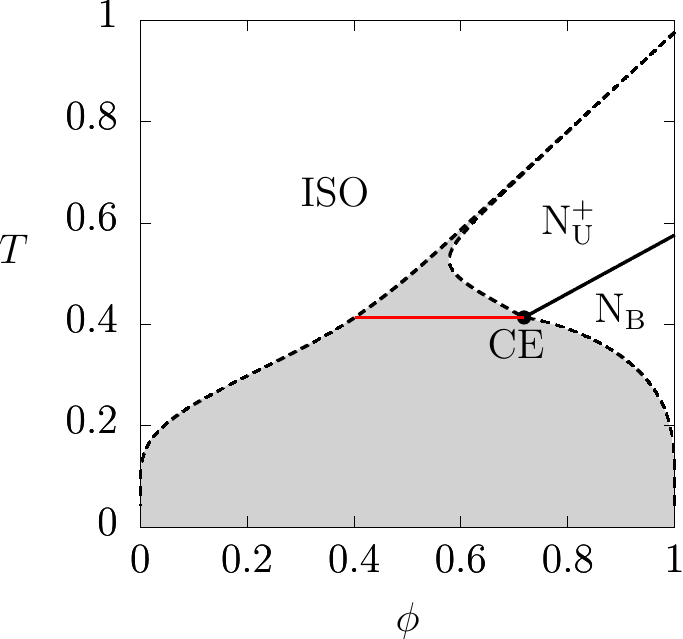}
      \caption{$\Delta=19/20$}
      \label{fig:2}
    \end{subfigure}
    \begin{subfigure}[b]{0.45\textwidth}
      \includegraphics[width=\textwidth]{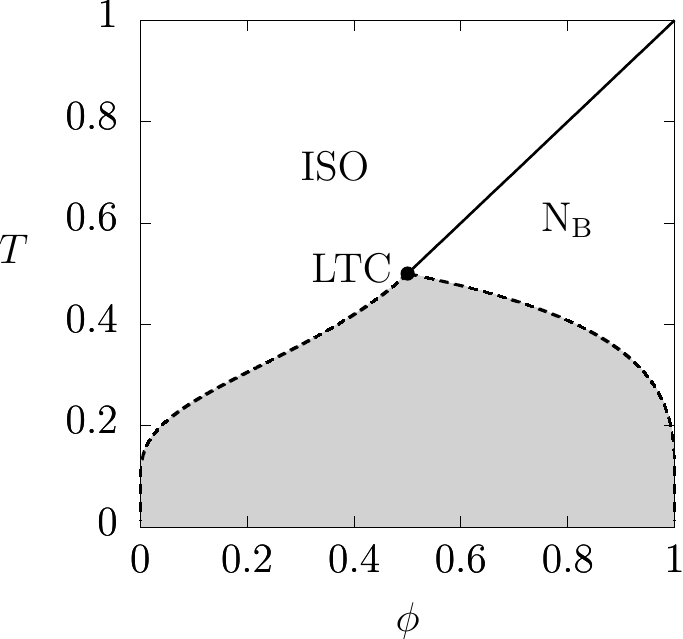}
      \caption{$\Delta=1$}
      \label{fig:3}
    \end{subfigure}
    \caption{Phase diagram in terms of temperature $T$ (in units of $A$) and concentration $\phi$ of nematogens, for different values of biaxiality degree  and in the absence of isotropic interactions ($U=0$). ISO: isotropic phase. N$_{\mathrm{U}}^{+}$: calamitic uniaxial nematic phase. N$_\mathrm{B}$: biaxial nematic phase. Dashed lines are the boundaries of biphasic region (grey). Red solid line: critical end point ($\mathrm{CE}$). LTC is a Landau tricritical point.}
\end{figure}

By considering intrinsically rod-like nematogens, for which $\Delta=0$, we find the phase diagram shown in Fig.\,\ref{fig:1}, which is qualitatively equivalent and quantitatively similar to the one obtained by Monte Carlo simulations for the 
Lebwohl--Lasher lattice-gas model of Ref.\,\cite{bates2001computer}, in the absence of isotropic interactions.
At high concentration ($\phi\gtrsim 0.75$), as $T$ decreases, the observed phase sequence is isotropic (ISO), followed by a biphasic region of coexisting rod-rich uniaxial nematic (N$_\mathrm{U}^{+}$) and rod-poor isotropic phases, followed by a pure uniaxial nematic and finally a reentrant coexistence region. At lower rod concentration the coexistence region is stable at low temperatures. 
The coexistence lines signaling the discontinuous transition from the isotropic phase to the uniaxial nematic phase is determined by Eq.\,(\ref{eq:mf}) evaluated at $\left(S,\eta,\phi\right)=\left( S_\mathrm{U}, 0, \phi_\mathrm{U} \right)$ and at $\left(S,\eta,\phi\right)=\left( 0, 0, \phi_\mathrm{I} \right)$, supplemented by $\psi(S_{\mathrm{U}},0,\phi_{\mathrm{U}})=\psi(0,0,\phi_{\mathrm{I}})$,
%
%
where $\phi_{\mathrm{I}}$ and $\phi_{\mathrm{U}}$ are respectively the concentrations of the isotropic and uniaxial phases at the transition point, and $S_{\mathrm{U}}$ is the value of $S$ at that point. Notice that, since the nematogens are intrinsically uniaxial, we can assume $\eta = 0$ without loss of generality. 


For the case of objects that are non-cylindrical, $\Delta \neq 0$ and $\Delta \neq 3$, it is possible to observe stable biaxial phases (N$_\mathrm{B}$), as shown in  Fig.\,\ref{fig:2} for biaxiality degree $\Delta=19/20$.
%
%
In this diagram, at high concentrations and high temperatures, there is a small biphasic region of coexisting uniaxial and isotropic phases. As temperature decreases, we have a second-order transition from the N$_{\mathrm{U}}^{+}$ phase to a pure N$_\mathrm{B}$ phase, and finally the biphasic region ISO-N$_\mathrm{B}$ appears. The conditions for determining the first-order transition between ISO and N$_\mathrm{B}$ are given by Eq.\,(\ref{eq:mf}) evaluated at $\left(S,\eta,\phi\right)=\left( S_\mathrm{B}, \eta_\mathrm{B}, \phi_\mathrm{B} \right)$ and at $\left(S,\eta,\phi\right)=\left( 0, 0, \phi_\mathrm{I} \right)$, as well as $\psi\left( S_\mathrm{B}, \eta_\mathrm{B}, \phi_\mathrm{B} \right) = \psi\left( 0, 0, \phi_\mathrm{I} \right) $, 
where $S_{\mathrm{B}}$ and $\eta_{\mathrm{B}}$ are the values taken by $S$ and $\eta$ in the biaxial state at the transition point. On the other hand, the second-order transition between uniaxial and biaxial phases is located by Eq.\,(\ref{eq:mf}) and $\partial^2 \psi / \partial \eta^2 = 0$, all evaluated at $\left(S,\eta,\phi\right)=\left( S_o, 0 , \phi_o\right)$, where $S_o$ and $\phi_o$ are the values of $S$ and $\phi$ at the transition point.
%
%
We also find that the  N$^{+}_{\mathrm{U}}$-N$_{\mathrm{B}}$ transition line meets the ISO-N$_{\mathrm{B}}$ biphase region at a critical end point ($\mathrm{CE}$), in which a critical nematic state separating uniaxial and biaxial phases coexists with a noncritical isotropic state. Critical end points are possible multicritical points that can be found in thermodynamics systems with many components \cite{Uzunov_Book,Mario_Book}. In our case, we have critical end points related to nematic transitions in a lattice-gas model with orientation-dependent interactions. These kinds of multicritical points were also reported in a Maier--Saupe model that mimics binary mixtures of uniaxial and biaxial nematogens \cite{nascimento2015maier}. 

For anisotropic objects with maximal biaxiality degree, $\Delta = 1$, stable uniaxial phases are absent and the phase diagrams present the general aspect shown in Fig.\,\ref{fig:3}. 
In this case, for high temperatures and high concentration, the $\mathrm{ISO}$-$\mathrm{N_{B}}$ transition is continuous and determined by the conditions $\partial \psi / \partial \phi = \partial \psi / \partial \eta = \partial^2 \psi / \partial \eta^2 = 0$, evaluated at the transition point $\left(S,\eta,\phi\right)=\left(0,0,\phi_o \right)$.
%
%
This line of continuous transitions is actually a line of multicritical Landau points. On the other hand, for low $T$ and intermediate concentrations, we observe an ISO-N$_{\mathrm{B}}$ coexistence region associated with a first-order transition at which $\partial \psi / \partial \eta = \partial \psi / \partial \phi = 0$ at $\left(S,\eta,\phi\right)=\left( 0, \eta_\mathrm{B},\phi_\mathrm{B}\right)$, $\partial \psi / \partial \phi =0$ at $\left(S,\eta,\phi\right)=\left( 0, 0,\phi_\mathrm{I}\right)$, and $\psi(0,\eta_{\mathrm{B}},\phi_{\mathrm{B}})=\psi(0,0,\phi_{\mathrm{I}})$.
%
%
The discontinuous and continuous transitions meet at a multicritical point which we call Landau tricritical (LTC) point. Roughly speaking, according to the solutions of mean-field equations, the multicritical point LTC has properties common to both Landau points \cite{GRAMSBERGEN1986} and tricritical points \cite{Uzunov_Book,Mario_Book}. Notice that in the limit of a pure system (i.e. $\phi \to 1$) consisting of biaxial objects with $\Delta =1$, our findings are in agreement with earlier mean-field results, which shown a direct ISO-N$_\mathrm{B}$ transition through a single, isolated Landau point in the $\Delta$-$T$ phase diagram \cite{nascimento2015maier,nascimento2016lattice}.

It is possible to determine the conditions that characterize a Landau tricritical point by following the discussion presented by Rodrigues \textit{et al.} \cite{rodrigues2020magnetic}. Indeed, in our context, an LTC point 
is the endpoint of a line of Landau points, and a Landau point happens when the stable solutions of MF equations for ISO and N$_{\mathrm{U}}^{\pm}$ become degenerate. Each point on a Landau line satisfies $\partial \psi / \partial \phi =  d^2 \psi / d S^2 = d^3 \psi / d S^3 = 0$, evaluated at $\left(S,\eta,\phi\right)=\left( 0,0,\phi_L\right)$.
%
%
Observe that these conditions involve partial derivatives as well as total derivatives (with respect to $S$) of the free-energy functional $\psi$. We must treat $\phi$ as an implicit function of $S$ while calculating the total derivatives. Thus, one can find  $\Delta=1$, $(\beta A-1)e^{\beta\mu}-1=0$ and $ \beta A  \phi_\mathrm{L} = 1 $, which are the same results obtained in Ref. \cite{rodrigues2020magnetic}. The solutions to these equations define a line of Landau points, which is represented by a solid line in Fig.\,\ref{fig:3}. In the limiting case of maximum concentration of biaxial objects, i.e. $\beta\mu\gg1$ or equivalently $\phi_{\mathrm{L}}\rightarrow1$, we recover the results obtained in previous treatments \cite{nascimento2015maier,nascimento2016lattice}, apart from differences in the definitions of parameters.
Nevertheless, we also have to check whether the solution leading to a Landau point corresponds to a minimum of the free-energy functional. This can be done by analyzing the behavior of the total fourth-order derivative of $\psi$ with respect to $\eta$ at $\left(S,\eta,\phi\right)=\left( 0,0,\phi_ \mathrm{L}\right)$, which gives $d^4 \psi / d  \eta^4 = -3 A^4 \beta^3\phi_\mathrm{L} \left( 1 - 2\phi_\mathrm{L} \right)/8 $.
%
%
This total derivative should be positive for stable states, but we notice that it may change its sign from positive, for $\phi_{\mathrm{L}}>1/2$, to negative, for $\phi_{\mathrm{L}}<1/2$, indicating that the Landau point is stable only if $\phi_{\mathrm{L}}>1/2$ (implying $\beta A<2$). Thus, precisely at $\phi_{\mathrm{L}}=1/2$, both $d^2 \psi / d  \eta^2$ and $d^4 \psi / d  \eta^4$ are zero, setting the conditions for locating a tricritical point that is also a Landau point. The coordinates of the LTC point are $\left( \beta A \right)_\mathrm{LTC}=2$, $\phi_{\mathrm{LTC}}=1/2$, and  $\mu_{\mathrm{LTC}}=0$. The stability of the LTC point can be checked by looking at the sixth-order derivative of $\psi$ with respect to $\eta$, which gives $d^{6}\psi/\,d\eta^{6}=2A >0 $, therefore corresponding to a free-energy minimum.

\begin{figure}[ht]
    \includegraphics[width=0.6\textwidth]{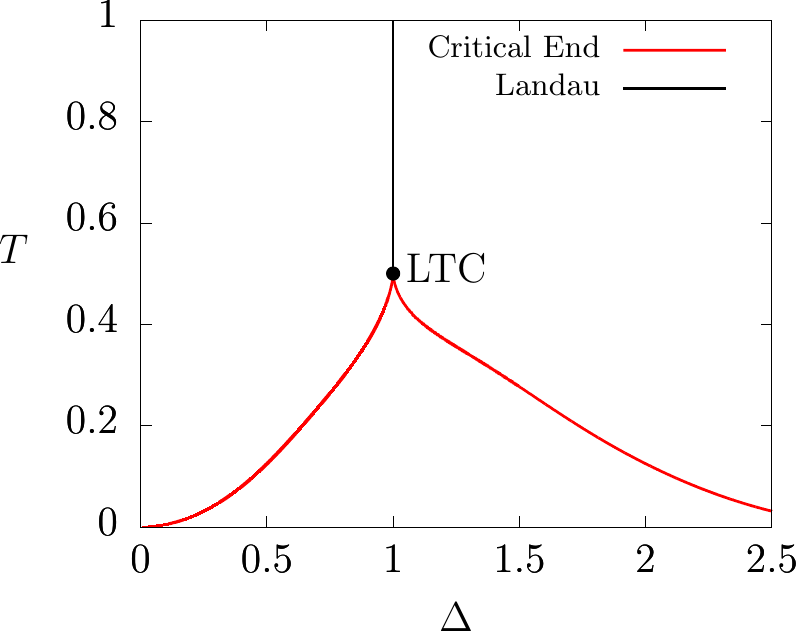}
    \caption{Lines of multicritical points in the plane $\Delta$-$T$ for zero isotropic interaction. The line of Landau points (black) meets the lines of critical end points (red) at a Landau tricritical (LTC) point, which is only present for maximal biaxiality degree $\Delta=1$.}
    \label{fig:4}
\end{figure}

We plotted all the lines of multicritical points obtained until now in the $\Delta$-$T$ plane shown in Fig.\,\ref{fig:4}. It is worth mentioning that, as we are assuming zero isotropic interaction, the space of thermodynamic fields is spanned by temperature $T$, chemical potential $\mu$ and biaxiality $\Delta$. Due to that, the lines presented in Fig.\,\ref{fig:4} are critical solutions of MF equations with varying chemical potential. Besides, although we have focused the discussion on calamitic nematic phases, for which $0<\Delta<1$, the results for discotic nematics (see e.g. Ref. \cite{Luders2021}) with $1<\Delta<3$ lead to phase diagrams with analogous topologies. Observe that for systems with maximal biaxiality degree, the LTC point occurs when the line of Landau points meets the two lines of critical end points. The Landau tricritical point is present only for maximal biaxiality $\Delta = 1$.

\section{Behavior in the presence of the  isotropic interaction}\label{sec4}

We now discuss phase diagrams in the presence of an isotropic interaction $U\neq 0$. In addition to uniaxial and biaxial structures, we may observe coexistence between isotropic fluid-like phases, which we call isotropic liquid ($\mathrm{IL}$) and isotropic vapor ($\mathrm{IV}$), as well as between nematic phases with different nematogen concentrations.

\subsection{Phase diagrams for uniaxial prolate nematogens (\texorpdfstring{$\Delta=0$}{}) }\label{subsecA}

\begin{figure}[ht]
    \begin{subfigure}[b]{0.45\textwidth}
     \includegraphics[width=\textwidth]{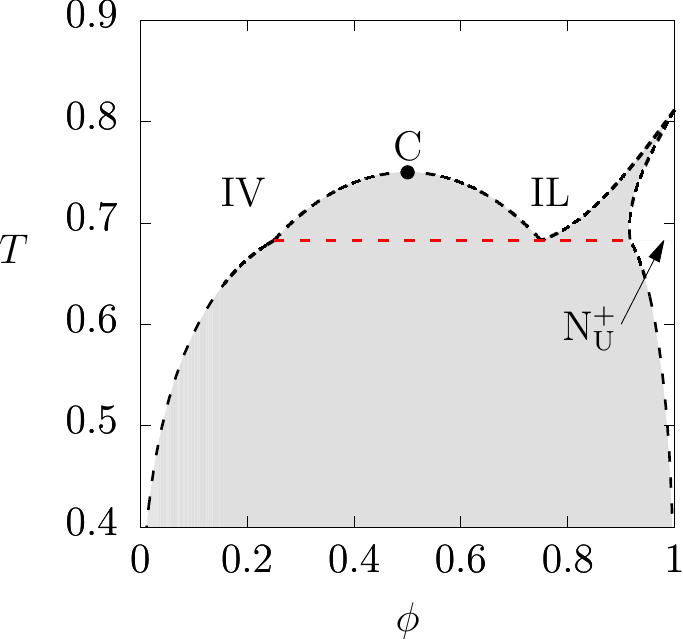}
     \caption{$(A, U)=(1,-3)$}
     \label{fig:5}
    \end{subfigure}\hfill
    \begin{subfigure}[b]{0.45\textwidth}
     \includegraphics[width=\textwidth]{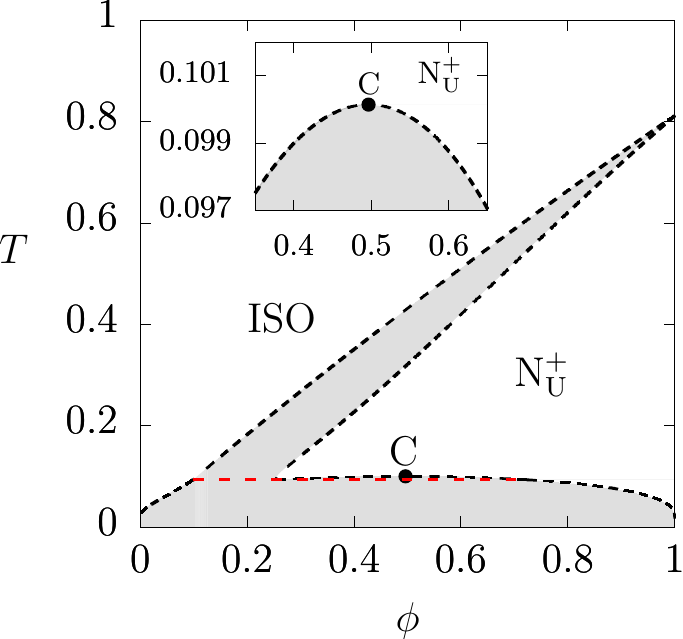}
     \caption{$(A,U)=(1,11/10)$}
     \label{fig:6}
    \end{subfigure}
    \caption{Phase diagrams in terms of temperature $T$ and concentration $\phi$ of nematogens, for an intrinsically uniaxial system ($\Delta=0$). Red dashed line: triple point. Black dashed line: first-order transitions. $\mathrm{C}$ is a simple critical point.}
\end{figure}

For intrinsically uniaxial, rod-like objects, a sufficiently attractive ($U<0$) isotropic interaction leads to the appearance of a vapor-liquid (or a  high-density--low-density transition) coexistence analogous to the van der Walls condensation, see Fig.\,\ref{fig:5}. 
The vapor-liquid transition is determined by $\partial \psi / \partial \phi = 0$ at $\left(S,\eta,\phi\right)=(0,0,\phi_{\mathrm{IV}})$ and at $\left(S,\eta,\phi\right)=(0,0,\phi_{\mathrm{IL}})$, in addition to $\psi(0,0,\phi_{\mathrm{IV}})=\psi(0,0,\phi_{\mathrm{IL}})$.
%
%
These first-order lines meet at a simple critical point ($\mathrm{C}$), located at 
$\phi_{\mathrm{C}}=1/2$, $\beta_{\mathrm{C}}=-4/U$, $\mu_{\mathrm{C}}=U/2$ with $\psi_{\mathrm{C}}=U[2\ln{(12)}-1]/8$.

%

We also find a vapor-liquid-uniaxial triple point, which is determined by evaluating Eq.\,\eqref{eq:mf} at $\left(S,\eta,\phi\right)=\left(0,0,\phi_\mathrm{IV}\right)$, at $\left(S,\eta,\phi\right)=\left(0,0,\phi_\mathrm{IL}\right)$ and at $\left(S,\eta,\phi\right)=\left(S_\mathrm{U},0,\phi_\mathrm{U}\right)$, in addition to imposing $\psi\left(0,0,\phi_\mathrm{IV}\right)=\psi\left(0,0,\phi_\mathrm{IL}\right)=\psi\left(S_\mathrm{U},0,\phi_\mathrm{U}\right)$. 
%
%
For $T$ values lower than the triple-point temperature, the IV-IL discontinuous transition becomes metastable with respect to the IV-N$_{\mathrm{U}}^{+}$ first-order transition. As the strength $|U|$ of the attractive interaction increases, the region of stability of N$_{\mathrm{U}}^{+}$ decreases and tends to become limited to a very small region near $\phi=1$; see Fig.\,\ref{fig:5}.
%
This reduction in the area of the uniaxial phase was observed by Bates, using Monte Carlo simulations, in a lattice-gas extension of the Lebwohl--Lasher model \cite{bates2001computer} and later in the model proposed by Humphries, Luckhurst and Romano \cite{bates2002phase}.
%

For repulsive isotropic interactions ($U>0$), it is possible to notice the appearance of a very narrow coexistence region between uniaxial nematic phases, as shown in Fig.\,\ref{fig:6}. This biphasic coexistence region between uniaxial structures presents an ordinary critical point C, which can be found by imposing the conditions $\partial \psi / \partial S = \partial \psi / \partial \phi = d^2 \psi / d \phi^2 = d^3 \psi / d \phi^3 = 0$, evaluated at $\left(S,\eta,\phi \right) = \left( S_\mathrm{C}, 0, \phi_\mathrm{C} \right)$.
%

\begin{figure}[ht]
    \includegraphics[width=0.6\textwidth]{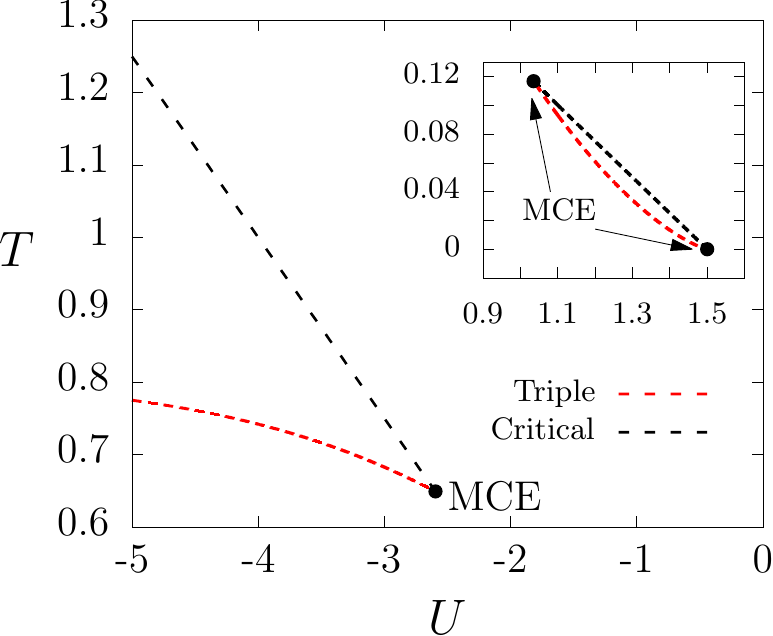}
    \caption{Lines of critical and of triple points in the $U$-$T$ plane, for the case of rod-like nematogens ($\Delta=0$). We notice that the lines of critical points meet the lines of triple points at higher-order multicritical end points ($\mathrm{MCE}$). The inset shows the case for repulsive isotropic interaction ($U>0$).}
    \label{fig:7}
\end{figure}

We plot the lines of critical points and of triple points in the $U$-$T$ plane in Fig.\,\ref{fig:7}. These lines meet at higher-order critical points, which we call multicritical end points (MCE), in analogy with critical end points appearing when lines of first-order and second-order transitions meet. For  $U<U^{(1)}_{\mathrm{MCE}}\approx-2.596$, we find phase diagrams with a simple critical point related to an IV-IL biphase region, in addition to a vapor-liquid-uniaxial triple point. This kind of phase phenomenon is associated with an attractive character of the isotropic interaction.
%
%
Nevertheless, for $U>U^{(1)}_{\mathrm{MCE}}$, it is no longer possible to distinguish between the IV and IL phases, and from a thermodynamic perspective there is a single isotropic phase. Then, we have phase diagrams which only show ISO-N$_\mathrm{U}$ coexistence regions. 

In the case of repulsive isotropic interactions with $U<U^{(2)}_{\mathrm{MCE}}\approx1.035$, the phase diagrams also exhibit first-order transitions between isotropic and uniaxial phases. However, for  $U^{(2)}_{\mathrm{MCE}}<U<U^{(3)}_{\mathrm{MCE}}=3/2$, as illustrated in Fig.\,\ref{fig:6}, it is possible to find phase diagrams exhibiting a coexistence region between uniaxial structures, with an associated critical point, as well as a triple point connecting one isotropic and two uniaxial states. As $U$ increases, we notice a decrease in the area of the low-temperature isotropic-uniaxial coexistence region, together with the decrease in the temperature of the critical and the triple points, until the ISO-N$_\mathrm{U}$ coexistence disappears completely as $U\to U^{(3) }_{\mathrm{MCE}}$. For this limiting value of $U$, both the temperatures of the critical point and of the triple point become zero.

\subsection{Phase diagrams for \texorpdfstring{$0<\Delta<1$}{}}

\begin{figure}[ht]
   \begin{subfigure}[b]{0.45\textwidth}
      \includegraphics[width=\textwidth]{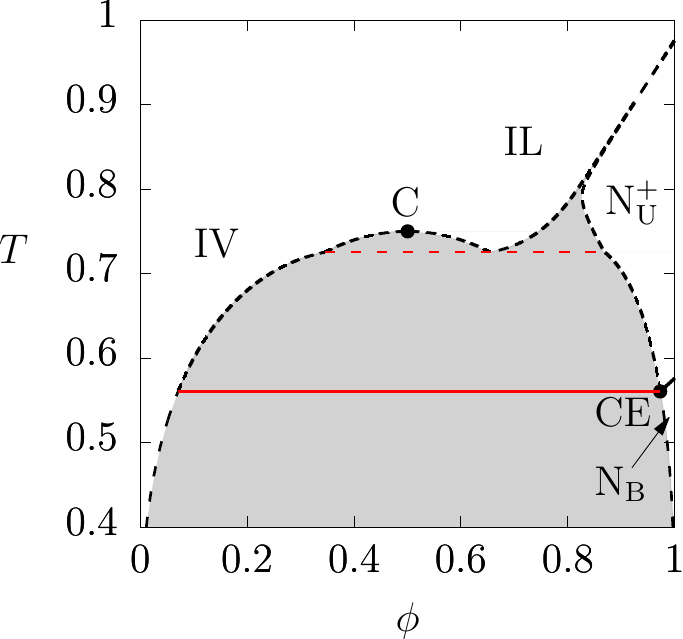}
      \subcaption{$\Delta=19/20$ and $(A, U)=(1,-3)$}
      \label{fig:66}
   \end{subfigure}\hfill
   \begin{subfigure}[b]{0.45\textwidth}
      \includegraphics[width=\textwidth]{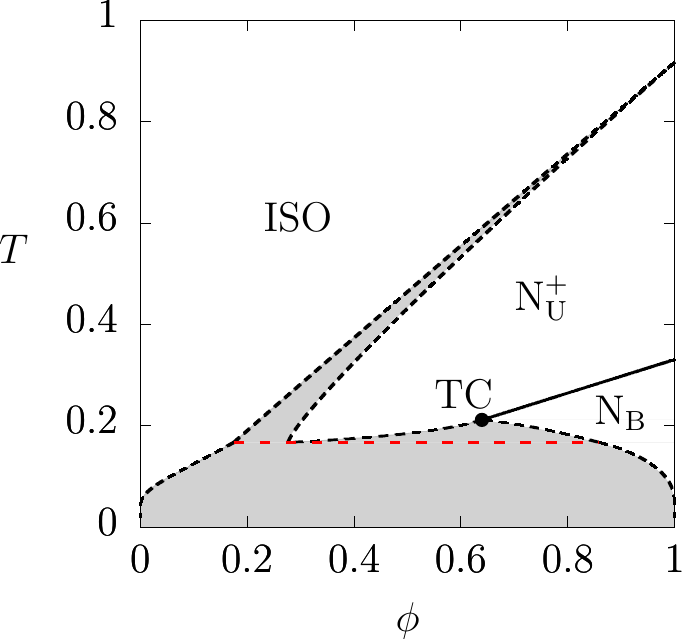}
      \subcaption{$\Delta=4/5$ and $(A, U)=(1,1)$}
      \label{fig:8}
   \end{subfigure}
   \begin{subfigure}[b]{0.45\textwidth}
      \includegraphics[width=\textwidth]{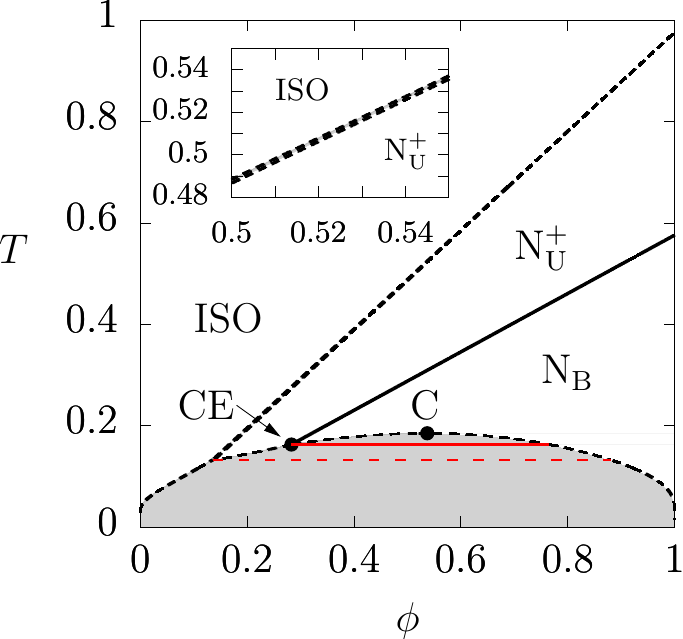}
      \subcaption{$\Delta=9/20$ and $(A, U)=(1,13/10)$}
      \label{fig:9}
   \end{subfigure}
   \caption{Phase diagrams in terms of temperature $T$ (in units of $A$) and concentration $\phi$ of nematogens, for different values of the biaxiality degree. The red dashed line represents a triple point. The red solid line is associated with a critical end point ($\mathrm{CE}$), while $\mathrm{C}$ is a critical point and $\mathrm{TC}$ is a tricritical point. The inset shows the isotropic-uniaxial coexistence region.}
\end{figure}

As previously mentioned, the discrete-state Maier--Saupe model presents phase diagrams with stable biaxial structures when the nematogens are intrinsically biaxial \cite{nascimento2015maier,nascimento2016lattice}. Then, we expect that the presence of dilution and isotropic interactions may lead to phase diagrams with more elaborate topologies. Indeed, for systems with attractive isotropic interactions, we obtain phase diagrams of the type shown in Fig.\,\ref{fig:66}. In this case, we have a critical point $\mathrm{C}$ associated with an IV-IL biphasic region, and an IV-IL-N$_\mathrm{U}^{+}$ triple point, analogous to those discussed in Sec.\,\ref{subsecA} for intrinsically uniaxial nematogens. We also find an IV-N$_\mathrm{B}$ discontinuous transition, determined by the conditions in Eq.\,\eqref{eq:mf}, evaluated at $\left(S,\eta,\phi\right)=\left( 0,0, \phi_\mathrm{IV}\right)$ and at $\left(S,\eta,\phi\right)=\left( S_\mathrm{B}, \eta_\mathrm{B}, \phi_\mathrm{B}\right)$, supplemented by $\psi(0,0,\phi_{\mathrm{IV}})=\psi(S_{\mathrm{B}},\eta_{\mathrm{B}},\phi_{\mathrm{B}})$. The coexistence between the biaxial phase and the isotropic vapor is verified at low temperatures, below the temperature of a critical end point CE ($T_{\mathrm{CE}}\approx0.56$ in the figure), whose location is set by Eq.\,\eqref{eq:mf}, evaluated at $\left(S,\eta,\phi\right)=\left( 0,0, \phi_\mathrm{IV}\right)$ and at $\left(S,\eta,\phi\right)=\left( S_\mathrm{CE}, 0, \phi_\mathrm{CE}\right)$, supplemented by $\psi(0,0,\phi_{\mathrm{IV}})=\psi(S_{\mathrm{CE}},0,\phi_{\mathrm{CE}})$ and $\partial^2\psi/\partial\eta^2=0$ at $\left(S,\eta,\phi\right)=\left( S_\mathrm{CE}, 0, \phi_\mathrm{CE}\right)$.
%
%
The biaxial nematic phase is stable for high concentrations and small temperatures. 
%
%

Let us now consider repulsive isotropic interactions, $(A, U)=(1,1)$, with biaxiality degree $\Delta=4/5$. The phase diagram is shown in Fig.\,\ref{fig:8}, where it is possible to identify a triple point in which isotropic, uniaxial, and biaxial phases coexist, as well as a tricritical point (TC), which satisfies the conditions $\partial \psi / \partial \phi = \partial \psi / \partial S = d^2 \psi / d \eta^2 = d^4 \psi / d \eta^4 = 0$, evaluated at $\left(S,\eta,\phi\right)=\left( S_\mathrm{TC}, 0, \phi_\mathrm{TC}\right)$.
%
%
The total derivatives are determined by treating $S$ and $\phi$ as implicit functions of $\eta$. The boundaries of the coexistence region associated with uniaxial and biaxial phases are determined by Eq.\,\eqref{eq:mf} evaluated at $\left(S,\eta,\phi\right)=\left( S_\mathrm{U}, 0, \phi_\mathrm{U} \right)$ and at $\left(S,\eta,\phi\right)=\left( S_\mathrm{B}, \eta_\mathrm{B}, \phi_\mathrm{B} \right)$, as well as $\psi\left( S_\mathrm{U}, 0, \phi_\mathrm{U} \right) = \psi\left( S_\mathrm{B}, \eta_\mathrm{B} , \phi_\mathrm{B} \right)$.
%
%
%
We also show in Fig.\,\ref{fig:9} the phase diagram corresponding to the repulsive case with $\Delta=19/20$ and $(A,U)=(1,13/10)$. There are biphasic regions associated with ISO and N$_\mathrm{U}^{+}$, N$_\mathrm{U}^{+}$ and N$_\mathrm{B}$, and ISO and N$_\mathrm{B}$. Besides, there is a triple point marking the coexistence of ISO, N$_\mathrm{U}^{+}$ and N$_\mathrm{B}$. Finally, we observe the presence of a biaxial-biaxial coexistence region, whose boundaries are determined by Eq.\,\eqref{eq:mf} evaluated at $\left(S,\eta,\phi\right)=\left( S_1, \eta_1, \phi_1 \right)$ and at $\left(S,\eta,\phi\right)=\left( S_2, \eta_2, \phi_2 \right)$, supplemented by $\psi\left( S_1, \eta_1, \phi_1 \right)=\psi\left( S_2, \eta_2, \phi_2 \right)$.

\subsection{Phase diagrams for \texorpdfstring{$\Delta=1$}{}}\label{subsecC}

Following our discussion in Sec.\,\ref{sec3}, we can obtain the conditions leading to Landau points for the maximal biaxiality degree and investigate the possible presence of Landau tricritical points. Indeed, we find analogous features when nonzero isotropic interactions are considered. Nevertheless, the parameter $U$ plays an important role in the criteria for determining the LTC point. After performing the calculation, we find that the coordinates of the Landau point satisfy $(\beta A-1)e^{\beta \mu}=e^{U/A}$ and $\beta A \phi = 1$.
For $\mu\rightarrow\infty$, i.e. in the limit of a fully occupied lattice, we recover the expected phase diagram with $\beta A=1$ at the Landau point, whereas for $U/A\rightarrow0$, we obtain the results discussed in Sec.\,\ref{sec3}. As we already know, the stability of a Landau point is related to the existence of an absolute minimum of the free-energy functional, and high-order derivatives should be considered because we are dealing with a multicritical point. The fourth-order derivative is
\begin{equation}
\left. \dfrac{d^4\psi}{d\eta^4} \right|_{(0,0,\phi)} = -\dfrac{3}{8}A^{3}\beta^2\left[\dfrac{U+A^{2}\beta-A(2+\beta U)}{A^{2}\beta+U(\beta A-1)}\right],
\end{equation}
%
%
This fourth-order derivative changes sign when $A\left( A-U\right)\beta=2A-U$,
which sets the condition for a possible LTC point. Notice that, as long as the isotropic interaction is attractive ($U<0$), there is always a candidate Landau tricritical point (since $\beta$ must be positive). However, as in the case $U=0$, the stability of that point for $U\neq0$ must be checked by looking at the sixth-order derivative of $\psi$ with respect to $\eta$, 
\begin{equation}
   \left.\dfrac{\,d^{6}\psi}{\,d\eta^{6}}\right|_{(0,0,\phi)}=\dfrac{(U-2A)^{4}(8A^{2}-30AU+15U^{2})}{64A(A-U)^{4}}.
\end{equation}
We then note that, since $A>0$, any LTC points are locally unstable if the isotropic interaction is repulsive ($U>0$) and such that $0.32\lesssim U/A \lesssim 1.68$. 

\begin{figure}[ht]
   \begin{subfigure}[b]{0.45\textwidth}
      \includegraphics[width=\textwidth]{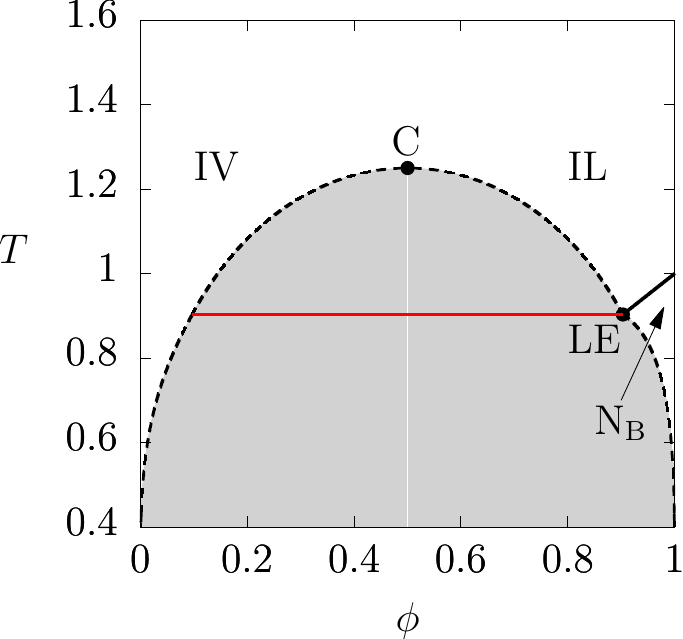}
      \subcaption{$(A,U)=(1,-5)$}
      \label{fig:11}
   \end{subfigure}\hfill
   \begin{subfigure}[b]{0.45\textwidth}
      \includegraphics[width=\textwidth]{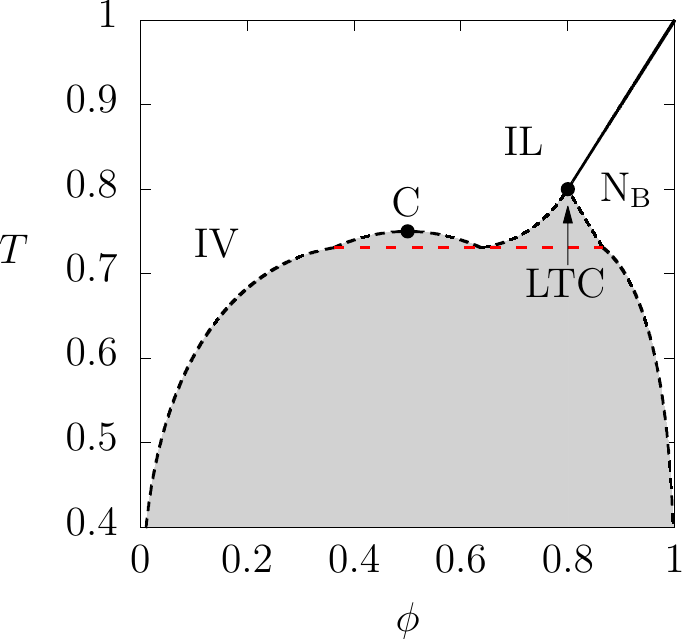}
      \subcaption{$(A,U)=(1,-3)$} 
      \label{fig:12}
   \end{subfigure}
   \begin{subfigure}[b]{0.45\textwidth}
      \includegraphics[width=\textwidth]{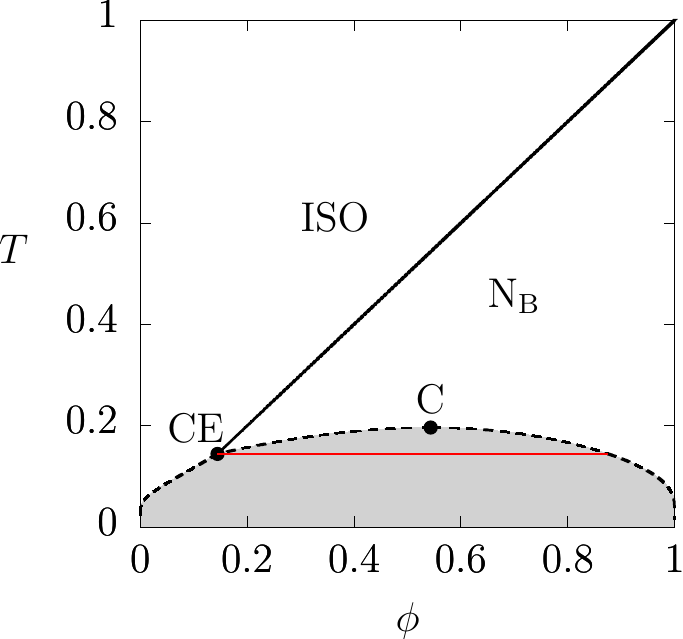}
      \subcaption{$(A,U)=(1,13/10)$}
      \label{fig:13}
   \end{subfigure}
   \caption{Phase diagram in terms of temperature $T$ (in units of $A$) and concentration $\phi$ of nematogens, for maximal biaxiality degree $\Delta=1$. The red dashed line represents a triple point. The red solid line represents a Landau critical end point (LE). $\mathrm{C}$ is a critical point.}
\end{figure}

For $U<0$, the LTC point is always locally stable, although it may not correspond to an absolute minimum of the free-energy functional. This is the case for $U=-5$, as shown by the phase diagram in Fig.\,\ref{fig:11}. There is a wide coexistence region associated with isotropic phases of vapor and liquid, and an ordinary critical point ($\mathrm{C}$). For high concentrations, as $T$ decreases, there exists a continuous transition from the IL phase to the N$_\mathrm{B}$ phase. Additionally, for a fixed sufficiently low temperature, by varying the concentration we enter a coexistence region between the IV and the N$_\mathrm{B}$ phases. The line of continuous transition consists of Landau points, and that line meets the coexistence regions at a Landau critical end point (LE). 
On the other hand, for isotropic interaction $U=-3$, we obtain the phase diagram exhibited in Fig.\,\ref{fig:12}. In this diagram, we now observe an LTC point, i.e. the LE point is not stable, and there also exists a triple point related to the IL, IV and N$_\mathrm{B}$ phases. When the isotropic interaction is sufficiently repulsive, we have a biaxial-biaxial coexistence region, as shown in Fig.\,\ref{fig:13}. This biphasic region presents a critical point C and a Landau critical end point LE. For phase diagrams with $U>2$, there are no coexistence regions and we only observe second-order transitions between the ISO and N$_\mathrm{B}$ phases; see Appendix \ref{appendix2}.

\begin{figure}[ht]
    \includegraphics[width=0.6\textwidth]{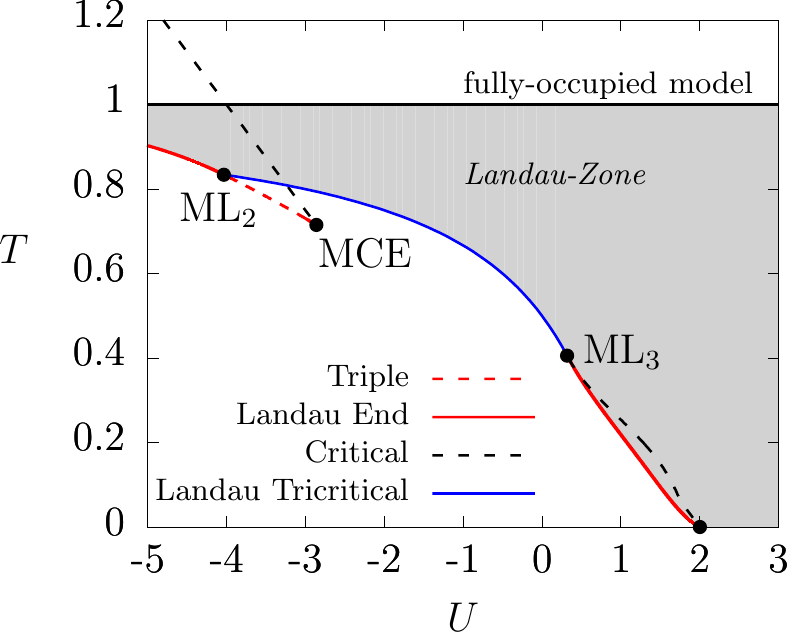}
    \caption{Lines of multicritical points in the $U$-$T$ plane for the case of maximal biaxiality parameter $\Delta=1$. The gray region marked as `Landau zone' consists of Landau points associated with different concentrations. ML$_2$ and ML$_3$ are higher-order Landau multicritical points. MCE is a higher-order multicritical end point.}
    \label{fig:14}
\end{figure}

For the particular case $\Delta=1$ we can plot a graph in the $U$-$T$ plane showing the multicritical points found for maximal biaxiality; see Fig.\,\ref{fig:14}. The corresponding phase diagrams in the $\phi$-$T$ plane present a line of Landau points regardless of the character of the isotropic interaction. The stability limits of points belonging to these Landau lines are (i) at high temperatures, the point $(\phi,T)=(1,1)$ (fully-occupied lattice) and (ii) at low temperatures, a multicritical point whose nature depends on the value of $U$. In the $U$-$T$ plane, the stable Landau points occupy an extensive region which we call the Landau zone. The boundaries of this region are the line $(\phi,T)=(1,1)$ and the lines of Landau critical end points and Landau tricritical points, which meet at multicritical Landau points ML$_2$ and ML$_3$. We also find a higher-order multicritical end point MCE related to a line of triple points. These triple points are associated with coexisting vapor, liquid and biaxial phases. Observe that the MCE point occurs when the line of triple points meets a line of critical points.

\section{Multicritical points in the biaxiality-temperature plane}

\label{sec5}
\begin{figure} 
    \begin{subfigure}[b]{0.48\textwidth}
     \includegraphics[width=\textwidth]{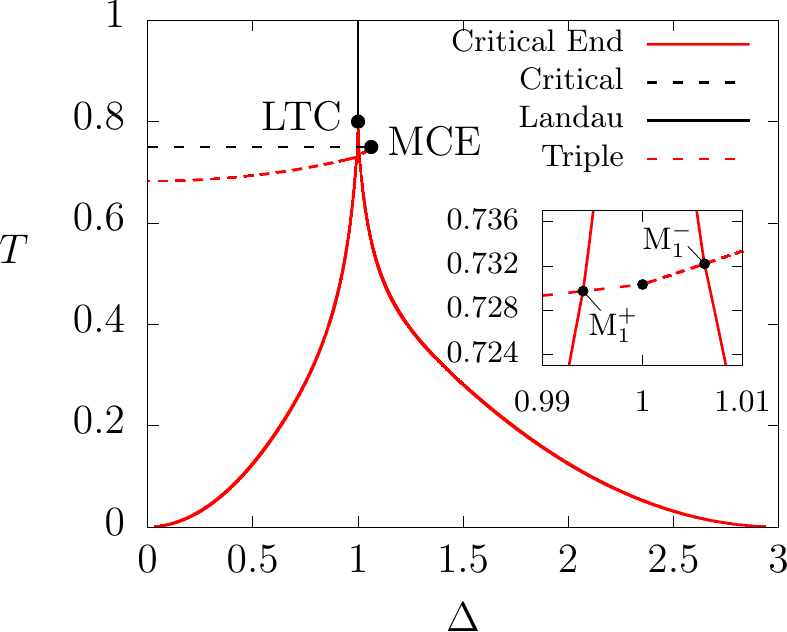}
     \subcaption{$U=-3$}
     \label{fig:16}
    \end{subfigure}
    \hfill
    \begin{subfigure}[b]{0.481\textwidth}
     \centering
     \includegraphics[width=\textwidth]{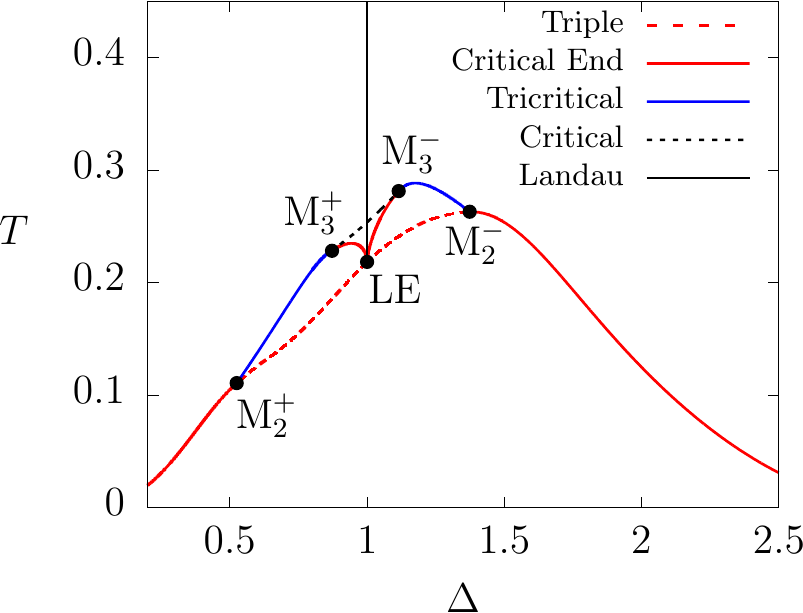}
     \subcaption{$U=1$} 
     \label{fig:15}
    \end{subfigure}
    \caption{Lines of multicritical points in the $\Delta$-$T$ plane. MCE: higher-order critical end point. M$^\pm_1$, M$^\pm_2$, and M$^\pm_3$ are higher-order multicritical points.}
    \label{figmulti}
\end{figure}

We may summarize the different topologies of the $\phi$-$T$ phase diagrams of the model by constructing diagrams of multicritical points in the plane $\Delta$-$T$ for a fixed value of $U$, as shown in Fig.\,\ref{figmulti}. Thus, given a nematic-like system with parameters $(A, U)$, we can determine the multicritical points in the $\phi$-$T$ phase diagrams for different values of $\Delta$. Due to the large parameter space, we only focus on some representative values of the isotropic interaction $U$.

\subsection{Case with $U<0$}

By assuming attractive isotropic interactions with $(A,U)=(1,-3)$, we obtain the $\Delta$-$T$ diagram shown in Fig.\,\ref{fig:16}. We notice that the line of triple points meets the lines of critical end points at higher-order multicritical points M$_1^{\pm}$. Besides, the line of ordinary critical points meets the line of triple points at the higher-order multicritical end point MCE. For  $\Delta<\Delta^{+}_1\approx0.994$, where $\Delta^{\pm}_1$ are the values of $\Delta$ at M$_1^{\pm}$, phase diagrams in the $\phi$-$T$ plane exhibit ordinary critical points related to vapor-liquid biphasic regions, critical end points (CE), and vapor-liquid-uniaxial triple points, a topology exemplified in Fig.\,\ref{fig:66}. Precisely at $\Delta=\Delta^{+}_1$, the lines of CE and triple points meet at the temperature $T^{+}_1\approx0.7298$. For values of model parameters corresponding to $\mathrm{M_{1}^\pm}$, $\phi$-$T$ phase diagrams do not exhibit a coexistence region between the isotropic vapor and the uniaxial phases. In the range $\Delta^{+}_1<\Delta<1$, the temperature of the CE point is higher than that of the triple point, which now represents a coexistence of isotropic (vapor and liquid) and biaxial phases. For maximal biaxiality $\Delta=1$, only isotropic and biaxial phases are stable, and $\phi$-$T$ phase diagrams are characterized by an ordinary vapor-liquid critical point, a Landau line, and, dependending on the value of $U<0$, a Landau tricritical point, as in Fig.\,\ref{fig:12}, or a Landau  end point, as in Fig.\,\ref{fig:11}. 

On the other hand, for $1<\Delta<\Delta^{-}_1\approx1.006$, the $\phi$-$T$ phase diagrams may exhibit uniaxial discotic phases, whose region of stability increases with $\Delta$. In addition, we have CE points and vapor-liquid-biaxial triple points, producing the same topology as in Fig.\,\ref{fig:66}. When $\Delta=\Delta^{-}_1$, the lines of $\mathrm{CE}$ and triple points meet at the temperature $T^{-}_1\approx0.7322$. For $\Delta^{-}_1<\Delta<\Delta_{\mathrm{MCE}}\approx1.063$, the $\phi$-$T$ phase diagrams also present $\mathrm{CE}$ points and vapor-liquid-biaxial triple points whose temperature approaches that of the vapor-liquid critical point as $\Delta\rightarrow\Delta_{\mathrm{MCE}}$. For biaxiality degree $\Delta=\Delta_{\mathrm{MCE}}$, the vapor-liquid-biaxial triple point and the ordinary vapor-liquid critical point meet at the temperature $T_{\mathrm{MCE}}=3/4$, and we cannot distinguish isotropic vapor and liquid phases. 
For nematic systems with $\Delta_{\mathrm{MCE}}<\Delta<3$, the topology of the $\phi$-$T$ phase diagrams is the same as the one shown in Fig.\,\ref{fig:2}, the only multicritical point being a CE point separating regions of coexistence between the low-concentration isotropic phase and the high-concentration biaxial (at low temperatures)  or uniaxial (at higher temperatures) phases.
For the intrinsically uniaxial cases $\Delta=0$ or $\Delta = 3$, the phase diagrams exhibit only isotropic and uniaxial phases, as exemplified in Sec.\,\ref{subsecA}.

\subsection{Case with $U>0$}

Now, by considering repulsive isotropic interactions with $(A,U)=(1,1)$, we obtain the multicritical lines shown in Fig.\,\ref{fig:15}. Here, lines of CE, tricritical and triple points meet at multicritical points M$_2^{\pm}$. We also have the multicritical points M$_3^{\pm}$, where lines of CE, triple and ordinary critical points meet. 
The topology of the $\phi$-$T$ phase diagrams is essentially symmetric with respect to the axis $\Delta=1$, except for the change in character of the uniaxial phases, from calamitic (for $0\leq\Delta<1$) to discotic (for $1<\Delta\leq 3$).

In the ranges $0<\Delta<\Delta^{+}_2\approx0.525$ or $\Delta_2^{-}\approx 1.3743 < \Delta < 3$, where $\Delta^{\pm}_{i}$ is the biaxiality parameter at $\mathrm{M}_i^{\pm}$, the topology of the $\phi$-$T$ phase diagrams is the same as the one shown in Fig.\,\ref{fig:2}, and the temperature of the CE point increases as the value of $\Delta$ becomes closer to $1$. For biaxiality in the ranges $\Delta^{+}_2<\Delta<\Delta^{+}_3\approx0.872$ or $\Delta^{-}_3\approx1.115 <\Delta<\Delta^{-}_2$, there exist isotropic-uniaxial and uniaxial-biaxial coexistence regions, as well as a tricritical (TC) point, as illustrated in Fig.\,\ref{fig:8}. Finally, for $\Delta^{+}_3 < \Delta < \Delta^{-}_3$, the TC point is replaced by a low-concentration CE point (or a Landau end point if $\Delta=1$) and an ordinary critical point associated with a biaxial-biaxial coexistence region, a topology exemplified in Fig.\,\ref{fig:13}. For biaxiality exactly equal to $\Delta^{+}_3$ or $\Delta^{-}_3$, the lines of critical and $\mathrm{CE}$ points meet the line of $\mathrm{TC}$ points and the biaxial-biaxial coexistence region is absent.

\section{Conclusions}\label{sec6}

We considered a lattice-gas version of the Maier--Saupe model for biaxial nematics with discrete orientations, in addition to an energetic term that described an isotropic interaction. The model is investigated in mean-field theory through a fully-connected spin-like system with inclusion of dilution effects. The free energy functional and the mean-field equations were obtained exactly. 

For systems without isotropic interactions, $U=0$, we have drawn phase diagrams in terms of temperature and concentration of nematogens, with fixed value of $\Delta$. The case $\Delta = 1$ is particularly interesting due to the absence of a nematic uniaxial phase, and we find a line of Landau points which is limited by a Landau tricritical point ($\mathrm{LTC}$). In the cases $\Delta=0$ or $3$ the nematogens are intrinsically uniaxial, so that the phase diagrams show no biaxial nematic phase. Any other value of $\Delta$ leads to a diagram which presents a critical end point ($\mathrm{CE}$) at high concentration. 

Systems with $U\neq0$ present a great variety of multicritical points depending on the character of the isotropic interaction and the biaxiality degree of the nematogens. To clarify this idea, diagrams of multicritical points were constructed in the $U$-$T$ plane for some values of $\Delta$, these diagrams show the different multicritical points that can be found in the phase diagrams. 

It would be interesting to extend the present work to deal with the limit in which the orientational interactions are described by the potential in Eq.\,\eqref{eq:V12} with $\zeta=0$ and $\lambda\neq0$. This would allow comparison with the results obtained by Skutnik \emph{et al.} \cite{Skutnik2020} for a three-dimensional model with short-range interactions via constant-pressure Monte Carlo simulations. Such a comparison would point to possible multicritical phenomena which could be further investigated via simulations.

Finally, we point out that our model could in principle be used to fit experimental data from lyotropic systems, providing estimates of coupling energies and biaxiality parameters, if we allow for variation of the parameter $\Delta$ with both temperature and concentration of components in a lyotropic mixture. Models for this variation should be informed by calculations similar to those provided by Amaral \emph{et al.} for the change in micelle form induced by cosurfactant addition \cite{Amaral1997}.

\section{Acknowledgment}
This work was funded by CNPq, FAPESP, INCT/FCx, NAP/FCx, and Coordenação de Aperfeiçoamento de Pessoal de Nível Superior – Brasil (CAPES) – Finance Code 001.

\appendix
\section{Mean-field calculations for LGMSZ model}\label{appendix1}
The mean-field version of the LGMSZ model is obtained
by assuming a fully-connected lattice Hamiltonian
\begin{equation}
    \mathcal{H}_\text{mf}=-\dfrac{A}{2N}\sum_{i,j}\gamma_{i}\gamma_{j}\mathbf{\Omega}_{i}\mathbf{:}\mathbf{\Omega}_{j}+\dfrac{U}{2N}\sum_{i,j}\gamma_{i}\gamma_{j},
\end{equation}
where the sums now run over all lattice sites. The grand partition function is 
\begin{equation}\label{eq:gpflgmszAp}
    \begin{split}
        \Xi=&\sum_{\{\gamma_{i}\}}\sum_{\{\mathbf{\Omega}_{i}\}}\exp\left(\dfrac{\beta A}{2N}\sum_{i,j}\gamma_{i}\gamma_{j}\mathbf{\Omega}_{i}\mathbf{:}\mathbf{\Omega}_{j}
        -\dfrac{\beta U}{2N}\sum_{i,j}\gamma_{i}\gamma_{j}+\beta\mu\sum_{i}\gamma_{i}\right).
    \end{split}
\end{equation}
In order to obtain an integral representation of the grand partition function in the mean-field limit, we introduce the concentration of nematogens as
\begin{equation}
    \phi=\dfrac{1}{N}\sum^{N}_{i=1}\gamma_{i},
\end{equation}
and use the integral representation of the Dirac delta function,
\begin{equation}\label{eq:deltaiden}
    \delta\left(N\phi-\sum^{N}_{i=1}\gamma_{i}\right)=\dfrac{1}{2\pi\im}\int^{+\im\infty}_{-\im\infty}\exp\left[-\hat{\phi}\left(N\phi-\sum^{N}_{i=1}\gamma_{i}\right)\right]\,d\hat{\phi},
\end{equation}
where $\im=\sqrt{-1}$ represents the imaginary unit. We also have the Gaussian identity
\begin{equation}\label{eq:id2}
        \exp\left(\dfrac{\beta A}{2N}\sum_{i,j}\gamma_{i}\gamma_{j}\mathbf{\Omega}_{i}\mathbf{:}\mathbf{\Omega}_{j}\right)\propto\int\exp\left(-\dfrac{\beta AN}{2}\norm{\mathbf{Q}}^{2}
        +\beta A\sum_{i}\gamma_{i}\mathbf{Q}\mathbf{:}\mathbf{\Omega}_{i}\right)\,d[\mathbf{Q}],
\end{equation}
where the constant of proportionality is irrelevant, and $\norm{\cdot}$ is the Frobenius norm. Using the identities in Eqs.\,\eqref{eq:deltaiden} and \eqref{eq:id2} and performing the partial trace over the occupation variables $\left\{\gamma_i\right\}$, we can write the grand partition function in the form 
\begin{equation}
    \Xi\propto\int\mathrm{I}(\mathbf{Q},\phi)e^{-N\beta\Gamma(\mathbf{
    Q},\phi)}\,d\phi\,d[\mathbf{Q}],
\end{equation}
where
\begin{equation}
    \Gamma(\mathbf{Q},\phi)=\dfrac{A}{2}\norm{\mathbf{Q}}^{2}+\dfrac{U}{2}\phi^{2}-\mu\phi,
\end{equation}
\begin{equation}\label{eq:iqp}
    \mathrm{I}(\mathbf{Q},\phi)=\dfrac{N}{2\pi\im}\int^{+\im\infty}_{-\im\infty}e^{Nf(\mathbf{Q},\phi,\hat{\phi})}\,d\hat{\phi},
\end{equation}
and
\begin{equation}\label{eq:f}
    f(\mathbf{Q},\phi,\hat{\phi})=-\hat{\phi}\phi+\ln\left(6+e^{\hat{\phi}}\sum_{\mathbf{\Omega}}e^{\beta A\mathbf{Q}\mathbf{:}\mathbf{\Omega}}\right).
\end{equation}
In the thermodynamic limit $N\gg1$, we expect the integral in Eq.\,\eqref{eq:iqp} to be dominated by the highest stationary point of $f(\mathbf{Q},\phi,\hat{\phi})$ with respect to $\hat{\phi}$. As for a complex function the only stationary points are saddle points, the integral is therefore dominated by the highest saddle point. The saddle point, $\hat{\phi}_{o}$, can be determined by the condition $f^{\prime}(\mathbf{Q},\phi,\hat{\phi}_{o})=0$, where the derivative is taken with respect to $\hat{\phi}$. Then
\begin{equation}
    \hat{\phi}_{o}=\ln\left(\dfrac{6\phi}{1-\phi}\right)+\ln\left(\sum_{\mathbf{\Omega}}e^{\beta A\mathbf{Q}\mathbf{:}\mathbf{\Omega}}\right),
\end{equation}
where $\hat{\phi}_{o}\in\mathbb{R}$, because $0<\phi<1$. In a neighborhood of $\hat{\phi}_{o}$ we can write 
\begin{equation}
    f(\mathbf{Q},\phi,\hat{\phi})\approx f(\mathbf{Q},\phi,\hat{\phi}_{o})+\dfrac{1}{2}f^{\prime\prime}(\mathbf{Q},\phi,\hat{\phi}_{o})(\hat{\phi}-\hat{\phi}_{o})^{2},
\end{equation}
so that the integral $\mathrm{I}(\mathbf{Q},\phi)$ takes the form
\begin{equation}\label{eq:approxI}
\begin{split}
    \mathrm{I}(\mathbf{Q},\phi)\approx&\dfrac{N}{2\pi\im}e^{N f(\mathbf{Q},\phi,\hat{\phi}_{o})}\\
    &\times\int^{+\im\infty}_{-\im\infty}\exp\left[\dfrac{N}{2}f^{\prime\prime}(\mathbf{Q},\phi,\hat{\phi}_{o})(\hat{\phi}-\hat{\phi}_{o})^{2}\right]\,d\hat{\phi}.
\end{split}
\end{equation}
The integral in Eq.\,\eqref{eq:approxI} can be solved by the method of steepest descents. For $\phi\approx\hat{\phi_o}$, we write
\begin{equation}
    \hat{\phi}-\hat{\phi}_{o}=\rho e^{\im\varphi},
\end{equation}
in which $\varphi$ is the angle according to which the integration contour passes through the saddle point $\hat{\phi}_{o}$ so that, in the complex plane defined by $\hat{\phi}$, $f^{\prime\prime}(\mathbf{Q},\phi,\hat{\phi_o})$ is a real number. 
Taking into account that in this particular problem $f^{\prime\prime}(\mathbf{Q},\phi,\hat{\phi_o})=\phi(1-\phi)$, implying $\varphi=\pi/2$ (see Ref.\,\cite{arfken2005}, p.\,491), we obtain 
\begin{equation}
    \mathbf{I}(\mathbf{Q},\phi)\approx\sqrt{\dfrac{N}{2\pi}}\dfrac{e^{N f(\mathbf{Q},\phi,\hat{\phi}_{o})}}{\sqrt{\phi(1-\phi)}}.
\end{equation}

Finally we get an integral representation of the grand partition function,
\begin{equation}
    \Xi\propto\int R(\phi)e^{-N\beta \psi(\phi,\mathbf{Q})}\,d\phi\,d[\mathbf{Q}],
\end{equation}
where
\begin{equation}\label{eq:app}
    \psi(\phi,\mathbf{Q})=\dfrac{A}{2}\norm{\mathbf{Q}}^{2}+\dfrac{U}{2}\phi^{2}-\mu\phi-\dfrac{f(\mathbf{Q},\phi,\hat{\phi}_{o})}{\beta},
\end{equation}
with
\begin{equation}
\begin{split}
   f(\mathbf{Q},\phi,\hat{\phi}_{o})&=-\phi\ln\phi-(1-\phi)\ln(\dfrac{1-\phi}{6})\\
   &\quad+\phi\ln\left[\sum_{\mathbf{\Omega}}\exp\left(\beta A\mathbf{Q}\mathbf{:}\mathbf{\Omega}\right)\right].
   \end{split}
\end{equation}

The symmetric traceless tensor $\mathbf{Q}$ can be parameterized by the scalar quantities $S$ and $\eta$ as
\begin{equation}
    \mathbf{Q}=\dfrac{1}{2}\begin{pmatrix}
 -S-\eta& 0 & 0\\ 
0 & -S+\eta & 0\\ 
0 & 0 & 2S 
\end{pmatrix},
\end{equation}
In terms of these parameters, the isotropic phase is characterized by $S=\eta=0$, the uniaxial phase by $S\neq0$ and $\eta=0$ (or $\eta=\pm3S$), and the biaxial phase by $\eta\neq0$. Using this parametrization in Eq.\,\eqref{eq:app}, we obtain the free-energy functional $\psi(S,\eta,\phi)$ in Eq.\,\eqref{eq:landaufef}.

\section{Low-temperature analysis}\label{appendix2}
Let us consider a diluted liquid crystal whose constituent units interact via the Hamiltonian of the LGMSZ model, which was presented in Sec.\,\ref{sec2}. Investigating the low-temperature limit $T\rightarrow0$ amounts to comparing the internal energy of the different phases, as minimizing this quantity for a given choice of the Hamiltonian parameters determines the stable phase. We must also consider the possibility that the internal energy is minimized under phase coexistence. 

In the isotropic phase, the energy is minimized by having $\mathbf{\Omega}_{i}\mathbf{:}\mathbf{\Omega}_{j}=0$ for any pair of particles $(i,j)$, so the internal energy as a function of $\phi$ is given by
\begin{equation}\label{eq:isoB}
    E_{\mathrm{I}}(\phi)=\dfrac{UN}{2}\phi^{2}.
\end{equation}
On the other hand, for $T\rightarrow 0$, $\mathbf{\Omega}_{i}\mathbf{:}\mathbf{\Omega}_{j}=(1+\Delta^{2})/2$ in the fully-occupied nematic phase (biaxial if $0<\Delta<3$ or uniaxial if $\Delta=0$ or $\Delta=3$) for any pair $(i,j)$. The internal energy of the nematic phase is
\begin{equation}
    E_{\mathrm{N}}(\phi)=-\dfrac{AN}{4}(3+\Delta^{2})\phi^{2}+\dfrac{UN}{2}\phi^{2}.
\end{equation}
As for the coexistence between an isotropic phase with $\phi=0$ and a nematic phase with $\phi\neq 0$, the lever rule gives an internal energy
\begin{equation}
    E_{\mathrm{I}\text{-}\mathrm{N}}=(1-\phi)E_{\mathrm{I}}(0)+\phi E_{\mathrm{N}}(1)=\phi E_{\mathrm{N}}(1).
\end{equation}

The sign of the energy difference $E_{\mathrm{I}\text{-}\mathrm{N}}-E_{\mathrm{N}}=\phi(1-\phi)E_{\mathrm{N}}(1)$ determines the stability of the nematic phase towards phase coexistence as $T\rightarrow0$. Just when $E_{\mathrm{N}}(1)=0$ the nematic phase becomes metastable with respect to isotropic-nematic coexistence. This corresponds to
\begin{equation}
    E_{\mathrm{N}}(1)=0\quad\Rightarrow\quad U=\dfrac{A}{2}(3+\Delta^{2}).
\end{equation}
Therefore, if $U/A>(3+\Delta^{2})/2$ the nematic phase is stable, otherwise there appears an isotropic-nematic coexistence region. 
\bibliographystyle{apsrev4-1}
\bibliography{references}

\begin{thebibliography}{38}%
\makeatletter
\providecommand \@ifxundefined [1]{%
 \@ifx{#1\undefined}
}%
\providecommand \@ifnum [1]{%
 \ifnum #1\expandafter \@firstoftwo
 \else \expandafter \@secondoftwo
 \fi
}%
\providecommand \@ifx [1]{%
 \ifx #1\expandafter \@firstoftwo
 \else \expandafter \@secondoftwo
 \fi
}%
\providecommand \natexlab [1]{#1}%
\providecommand \enquote  [1]{``#1''}%
\providecommand \bibnamefont  [1]{#1}%
\providecommand \bibfnamefont [1]{#1}%
\providecommand \citenamefont [1]{#1}%
\providecommand \href@noop [0]{\@secondoftwo}%
\providecommand \href [0]{\begingroup \@sanitize@url \@href}%
\providecommand \@href[1]{\@@startlink{#1}\@@href}%
\providecommand \@@href[1]{\endgroup#1\@@endlink}%
\providecommand \@sanitize@url [0]{\catcode `\\12\catcode `\$12\catcode
  `\&12\catcode `\#12\catcode `\^12\catcode `\_12\catcode `\%12\relax}%
\providecommand \@@startlink[1]{}%
\providecommand \@@endlink[0]{}%
\providecommand \url  [0]{\begingroup\@sanitize@url \@url }%
\providecommand \@url [1]{\endgroup\@href {#1}{\urlprefix }}%
\providecommand \urlprefix  [0]{URL }%
\providecommand \Eprint [0]{\href }%
\providecommand \doibase [0]{http://dx.doi.org/}%
\providecommand \selectlanguage [0]{\@gobble}%
\providecommand \bibinfo  [0]{\@secondoftwo}%
\providecommand \bibfield  [0]{\@secondoftwo}%
\providecommand \translation [1]{[#1]}%
\providecommand \BibitemOpen [0]{}%
\providecommand \bibitemStop [0]{}%
\providecommand \bibitemNoStop [0]{.\EOS\space}%
\providecommand \EOS [0]{\spacefactor3000\relax}%
\providecommand \BibitemShut  [1]{\csname bibitem#1\endcsname}%
\let\auto@bib@innerbib\@empty
\bibitem [{\citenamefont {de~Gennes}\ and\ \citenamefont
  {Prost}(1993)}]{deGennes_Book}%
  \BibitemOpen
  \bibfield  {author} {\bibinfo {author} {\bibfnamefont {P.}~\bibnamefont
  {de~Gennes}}\ and\ \bibinfo {author} {\bibfnamefont {J.}~\bibnamefont
  {Prost}},\ }\href {https://books.google.com.br/books?id=0Nw-dzWz5agC} {\emph
  {\bibinfo {title} {The Physics of Liquid Crystals}}},\ International Series
  of Monographs on Physics\ (\bibinfo  {publisher} {Clarendon Press},\ \bibinfo
  {year} {1993})\BibitemShut {NoStop}%
\bibitem [{\citenamefont {{Figueiredo Neto}}\ and\ \citenamefont
  {Salinas}(2005)}]{FigueiredoNeto2005}%
  \BibitemOpen
  \bibfield  {author} {\bibinfo {author} {\bibfnamefont {A.~M.}\ \bibnamefont
  {{Figueiredo Neto}}}\ and\ \bibinfo {author} {\bibfnamefont {S.~R.~A.}\
  \bibnamefont {Salinas}},\ }\href@noop {} {\emph {\bibinfo {title} {The
  physics of lyotropic liquid crystals}}}\ (\bibinfo  {publisher} {Oxford
  University Press},\ \bibinfo {year} {2005})\BibitemShut {NoStop}%
\bibitem [{\citenamefont {Singh}(2000)}]{Singh2000}%
  \BibitemOpen
  \bibfield  {author} {\bibinfo {author} {\bibfnamefont {S.}~\bibnamefont
  {Singh}},\ }\href {\doibase https://doi.org/10.1016/S0370-1573(99)00049-6}
  {\bibfield  {journal} {\bibinfo  {journal} {Physics Reports}\ }\textbf
  {\bibinfo {volume} {324}},\ \bibinfo {pages} {107} (\bibinfo {year}
  {2000})}\BibitemShut {NoStop}%
\bibitem [{\citenamefont {Palffy-Muhoray}(2007)}]{Palffy2007}%
  \BibitemOpen
  \bibfield  {author} {\bibinfo {author} {\bibfnamefont {P.}~\bibnamefont
  {Palffy-Muhoray}},\ }\href {\doibase 10.1063/1.2784685} {\bibfield  {journal}
  {\bibinfo  {journal} {Physics Today}\ }\textbf {\bibinfo {volume} {60}},\
  \bibinfo {pages} {54} (\bibinfo {year} {2007})}\BibitemShut {NoStop}%
\bibitem [{\citenamefont {Freiser}(1970)}]{freiser1970ordered}%
  \BibitemOpen
  \bibfield  {author} {\bibinfo {author} {\bibfnamefont {M.~J.}\ \bibnamefont
  {Freiser}},\ }\href {https://doi.org/10.1103/PhysRevLett.24.1041} {\bibfield
  {journal} {\bibinfo  {journal} {Physical Review Letters}\ }\textbf {\bibinfo
  {volume} {24}},\ \bibinfo {pages} {1041} (\bibinfo {year}
  {1970})}\BibitemShut {NoStop}%
\bibitem [{\citenamefont {Yu}\ and\ \citenamefont
  {Saupe}(1980)}]{yu1980observation}%
  \BibitemOpen
  \bibfield  {author} {\bibinfo {author} {\bibfnamefont {L.}~\bibnamefont
  {Yu}}\ and\ \bibinfo {author} {\bibfnamefont {A.}~\bibnamefont {Saupe}},\
  }\href {https://doi.org/10.1103/PhysRevLett.45.1000} {\bibfield  {journal}
  {\bibinfo  {journal} {Physical Review Letters}\ }\textbf {\bibinfo {volume}
  {45}},\ \bibinfo {pages} {1000} (\bibinfo {year} {1980})}\BibitemShut
  {NoStop}%
\bibitem [{\citenamefont {J\'akli}\ \emph {et~al.}(2018)\citenamefont
  {J\'akli}, \citenamefont {Lavrentovich},\ and\ \citenamefont
  {Selinger}}]{Jakli2018}%
  \BibitemOpen
  \bibfield  {author} {\bibinfo {author} {\bibfnamefont {A.}~\bibnamefont
  {J\'akli}}, \bibinfo {author} {\bibfnamefont {O.~D.}\ \bibnamefont
  {Lavrentovich}}, \ and\ \bibinfo {author} {\bibfnamefont {J.~V.}\
  \bibnamefont {Selinger}},\ }\href {\doibase 10.1103/RevModPhys.90.045004}
  {\bibfield  {journal} {\bibinfo  {journal} {Rev. Mod. Phys.}\ }\textbf
  {\bibinfo {volume} {90}},\ \bibinfo {pages} {045004} (\bibinfo {year}
  {2018})}\BibitemShut {NoStop}%
\bibitem [{\citenamefont {Luckhurst}\ and\ \citenamefont
  {Sluckin}(2015)}]{Luckhurst2015biaxial}%
  \BibitemOpen
  \bibfield  {author} {\bibinfo {author} {\bibfnamefont {G.}~\bibnamefont
  {Luckhurst}}\ and\ \bibinfo {author} {\bibfnamefont {T.}~\bibnamefont
  {Sluckin}},\ }\href {https://books.google.com.br/books?id=RhrPBwAAQBAJ}
  {\emph {\bibinfo {title} {Biaxial Nematic Liquid Crystals: Theory, Simulation
  and Experiment}}}\ (\bibinfo  {publisher} {Wiley},\ \bibinfo {year}
  {2015})\BibitemShut {NoStop}%
\bibitem [{\citenamefont {Akpinar}\ and\ \citenamefont {{Figueiredo
  Neto}}(2019)}]{Akpinar2019}%
  \BibitemOpen
  \bibfield  {author} {\bibinfo {author} {\bibfnamefont {E.}~\bibnamefont
  {Akpinar}}\ and\ \bibinfo {author} {\bibfnamefont {A.~M.}\ \bibnamefont
  {{Figueiredo Neto}}},\ }\href {\doibase 10.3390/cryst9030158} {\bibfield
  {journal} {\bibinfo  {journal} {Crystals}\ }\textbf {\bibinfo {volume} {9}}
  (\bibinfo {year} {2019}),\ 10.3390/cryst9030158}\BibitemShut {NoStop}%
\bibitem [{\citenamefont {Gramsbergen}\ \emph {et~al.}(1986)\citenamefont
  {Gramsbergen}, \citenamefont {Longa},\ and\ \citenamefont {{de
  Jeu}}}]{GRAMSBERGEN1986}%
  \BibitemOpen
  \bibfield  {author} {\bibinfo {author} {\bibfnamefont {E.~F.}\ \bibnamefont
  {Gramsbergen}}, \bibinfo {author} {\bibfnamefont {L.}~\bibnamefont {Longa}},
  \ and\ \bibinfo {author} {\bibfnamefont {W.~H.}\ \bibnamefont {{de Jeu}}},\
  }\href {\doibase https://doi.org/10.1016/0370-1573(86)90007-4} {\bibfield
  {journal} {\bibinfo  {journal} {Physics Reports}\ }\textbf {\bibinfo {volume}
  {135}},\ \bibinfo {pages} {195} (\bibinfo {year} {1986})}\BibitemShut
  {NoStop}%
\bibitem [{\citenamefont {Lebwohl}\ and\ \citenamefont
  {Lasher}(1972)}]{Lebwohl1972}%
  \BibitemOpen
  \bibfield  {author} {\bibinfo {author} {\bibfnamefont {P.~A.}\ \bibnamefont
  {Lebwohl}}\ and\ \bibinfo {author} {\bibfnamefont {G.}~\bibnamefont
  {Lasher}},\ }\href {\doibase 10.1103/PhysRevA.6.426} {\bibfield  {journal}
  {\bibinfo  {journal} {Phys. Rev. A}\ }\textbf {\bibinfo {volume} {6}},\
  \bibinfo {pages} {426} (\bibinfo {year} {1972})}\BibitemShut {NoStop}%
\bibitem [{\citenamefont {Bates}(2001)}]{bates2001computer}%
  \BibitemOpen
  \bibfield  {author} {\bibinfo {author} {\bibfnamefont {M.~A.}\ \bibnamefont
  {Bates}},\ }\href {https://doi.org/10.1103/PhysRevE.64.051702} {\bibfield
  {journal} {\bibinfo  {journal} {Physical Review E}\ }\textbf {\bibinfo
  {volume} {64}},\ \bibinfo {pages} {051702} (\bibinfo {year}
  {2001})}\BibitemShut {NoStop}%
\bibitem [{\citenamefont {Bates}(2002)}]{bates2002phase}%
  \BibitemOpen
  \bibfield  {author} {\bibinfo {author} {\bibfnamefont {M.~A.}\ \bibnamefont
  {Bates}},\ }\href {https://doi.org/10.1103/PhysRevE.65.041706} {\bibfield
  {journal} {\bibinfo  {journal} {Physical Review E}\ }\textbf {\bibinfo
  {volume} {65}},\ \bibinfo {pages} {041706} (\bibinfo {year}
  {2002})}\BibitemShut {NoStop}%
\bibitem [{\citenamefont {Luckhurst}\ and\ \citenamefont
  {Romano}(1980)}]{Luckhurst1980}%
  \BibitemOpen
  \bibfield  {author} {\bibinfo {author} {\bibfnamefont {G.}~\bibnamefont
  {Luckhurst}}\ and\ \bibinfo {author} {\bibfnamefont {S.}~\bibnamefont
  {Romano}},\ }\href {\doibase 10.1080/00268978000101341} {\bibfield  {journal}
  {\bibinfo  {journal} {Molecular Physics}\ }\textbf {\bibinfo {volume} {40}},\
  \bibinfo {pages} {129} (\bibinfo {year} {1980})}\BibitemShut {NoStop}%
\bibitem [{\citenamefont {Maier}\ and\ \citenamefont
  {Saupe}(1958)}]{maier1958einfache}%
  \BibitemOpen
  \bibfield  {author} {\bibinfo {author} {\bibfnamefont {W.}~\bibnamefont
  {Maier}}\ and\ \bibinfo {author} {\bibfnamefont {A.}~\bibnamefont {Saupe}},\
  }\href {https://doi.org/10.1515/zna-1958-0716} {\bibfield  {journal}
  {\bibinfo  {journal} {Zeitschrift f{\"u}r Naturforschung A}\ }\textbf
  {\bibinfo {volume} {13}},\ \bibinfo {pages} {564} (\bibinfo {year}
  {1958})}\BibitemShut {NoStop}%
\bibitem [{\citenamefont {Longa}\ and\ \citenamefont
  {Paj\c{a}k}(2005)}]{Longa2005}%
  \BibitemOpen
  \bibfield  {author} {\bibinfo {author} {\bibfnamefont {L.}~\bibnamefont
  {Longa}}\ and\ \bibinfo {author} {\bibfnamefont {G.}~\bibnamefont
  {Paj\c{a}k}},\ }\href {\doibase 10.1080/02678290500167873} {\bibfield
  {journal} {\bibinfo  {journal} {Liquid Crystals}\ }\textbf {\bibinfo {volume}
  {32}},\ \bibinfo {pages} {1409} (\bibinfo {year} {2005})}\BibitemShut
  {NoStop}%
\bibitem [{\citenamefont {Biscarini}\ \emph {et~al.}(1995)\citenamefont
  {Biscarini}, \citenamefont {Chiccoli}, \citenamefont {Pasini}, \citenamefont
  {Semeria},\ and\ \citenamefont {Zannoni}}]{Biscarini1995}%
  \BibitemOpen
  \bibfield  {author} {\bibinfo {author} {\bibfnamefont {F.}~\bibnamefont
  {Biscarini}}, \bibinfo {author} {\bibfnamefont {C.}~\bibnamefont {Chiccoli}},
  \bibinfo {author} {\bibfnamefont {P.}~\bibnamefont {Pasini}}, \bibinfo
  {author} {\bibfnamefont {F.}~\bibnamefont {Semeria}}, \ and\ \bibinfo
  {author} {\bibfnamefont {C.}~\bibnamefont {Zannoni}},\ }\href {\doibase
  10.1103/PhysRevLett.75.1803} {\bibfield  {journal} {\bibinfo  {journal}
  {Phys. Rev. Lett.}\ }\textbf {\bibinfo {volume} {75}},\ \bibinfo {pages}
  {1803} (\bibinfo {year} {1995})}\BibitemShut {NoStop}%
\bibitem [{\citenamefont {Straley}(1974)}]{Straley1974}%
  \BibitemOpen
  \bibfield  {author} {\bibinfo {author} {\bibfnamefont {J.~P.}\ \bibnamefont
  {Straley}},\ }\href {\doibase 10.1103/PhysRevA.10.1881} {\bibfield  {journal}
  {\bibinfo  {journal} {Phys. Rev. A}\ }\textbf {\bibinfo {volume} {10}},\
  \bibinfo {pages} {1881} (\bibinfo {year} {1974})}\BibitemShut {NoStop}%
\bibitem [{\citenamefont {Sonnet}\ \emph {et~al.}(2003)\citenamefont {Sonnet},
  \citenamefont {Virga},\ and\ \citenamefont {Durand}}]{sonnet2003dielectric}%
  \BibitemOpen
  \bibfield  {author} {\bibinfo {author} {\bibfnamefont {A.~M.}\ \bibnamefont
  {Sonnet}}, \bibinfo {author} {\bibfnamefont {E.~G.}\ \bibnamefont {Virga}}, \
  and\ \bibinfo {author} {\bibfnamefont {G.~E.}\ \bibnamefont {Durand}},\
  }\href {https://doi.org/10.1103/PhysRevE.67.061701} {\bibfield  {journal}
  {\bibinfo  {journal} {Physical Review E}\ }\textbf {\bibinfo {volume} {67}},\
  \bibinfo {pages} {061701} (\bibinfo {year} {2003})}\BibitemShut {NoStop}%
\bibitem [{\citenamefont {Horn}\ and\ \citenamefont
  {Johnson}(2013)}]{Horn_Book}%
  \BibitemOpen
  \bibfield  {author} {\bibinfo {author} {\bibfnamefont {R.~A.}\ \bibnamefont
  {Horn}}\ and\ \bibinfo {author} {\bibfnamefont {C.~R.}\ \bibnamefont
  {Johnson}},\ }\href@noop {} {\emph {\bibinfo {title} {Matrix Analysis}}},\
  \bibinfo {edition} {2nd}\ ed.\ (\bibinfo  {publisher} {Cambridge University
  Press},\ \bibinfo {year} {2013})\BibitemShut {NoStop}%
\bibitem [{\citenamefont {Nascimento}\ \emph {et~al.}(2015)\citenamefont
  {Nascimento}, \citenamefont {Henriques}, \citenamefont {Vieira},\ and\
  \citenamefont {Salinas}}]{nascimento2015maier}%
  \BibitemOpen
  \bibfield  {author} {\bibinfo {author} {\bibfnamefont {E.~S.}\ \bibnamefont
  {Nascimento}}, \bibinfo {author} {\bibfnamefont {E.~F.}\ \bibnamefont
  {Henriques}}, \bibinfo {author} {\bibfnamefont {A.~P.}\ \bibnamefont
  {Vieira}}, \ and\ \bibinfo {author} {\bibfnamefont {S.~R.}\ \bibnamefont
  {Salinas}},\ }\href {https://doi.org/10.1103/PhysRevE.92.062503} {\bibfield
  {journal} {\bibinfo  {journal} {Physical Review E}\ }\textbf {\bibinfo
  {volume} {92}},\ \bibinfo {pages} {062503} (\bibinfo {year}
  {2015})}\BibitemShut {NoStop}%
\bibitem [{\citenamefont {Zwanzig}(1963)}]{Zwanzig1963}%
  \BibitemOpen
  \bibfield  {author} {\bibinfo {author} {\bibfnamefont {R.}~\bibnamefont
  {Zwanzig}},\ }\href {\doibase 10.1063/1.1734518} {\bibfield  {journal}
  {\bibinfo  {journal} {The Journal of Chemical Physics}\ }\textbf {\bibinfo
  {volume} {39}},\ \bibinfo {pages} {1714} (\bibinfo {year}
  {1963})}\BibitemShut {NoStop}%
\bibitem [{\citenamefont {{de Oliveira}}\ and\ \citenamefont {{Figueiredo
  Neto}}(1986)}]{de1986reentrant}%
  \BibitemOpen
  \bibfield  {author} {\bibinfo {author} {\bibfnamefont {M.~J.}\ \bibnamefont
  {{de Oliveira}}}\ and\ \bibinfo {author} {\bibfnamefont {A.~M.}\ \bibnamefont
  {{Figueiredo Neto}}},\ }\href {https://doi.org/10.1103/PhysRevA.34.3481}
  {\bibfield  {journal} {\bibinfo  {journal} {Physical Review A}\ }\textbf
  {\bibinfo {volume} {34}},\ \bibinfo {pages} {3481} (\bibinfo {year}
  {1986})}\BibitemShut {NoStop}%
\bibitem [{\citenamefont {{do Carmo}}\ \emph {et~al.}(2010)\citenamefont {{do
  Carmo}}, \citenamefont {Liarte},\ and\ \citenamefont
  {Salinas}}]{do2010statistical}%
  \BibitemOpen
  \bibfield  {author} {\bibinfo {author} {\bibfnamefont {E.}~\bibnamefont {{do
  Carmo}}}, \bibinfo {author} {\bibfnamefont {D.~B.}\ \bibnamefont {Liarte}}, \
  and\ \bibinfo {author} {\bibfnamefont {S.~R.}\ \bibnamefont {Salinas}},\
  }\href {https://doi.org/10.1103/PhysRevE.81.062701} {\bibfield  {journal}
  {\bibinfo  {journal} {Physical Review E}\ }\textbf {\bibinfo {volume} {81}},\
  \bibinfo {pages} {062701} (\bibinfo {year} {2010})}\BibitemShut {NoStop}%
\bibitem [{\citenamefont {{do Carmo}}\ \emph {et~al.}(2011)\citenamefont {{do
  Carmo}}, \citenamefont {Vieira},\ and\ \citenamefont
  {Salinas}}]{do2011phase}%
  \BibitemOpen
  \bibfield  {author} {\bibinfo {author} {\bibfnamefont {E.}~\bibnamefont {{do
  Carmo}}}, \bibinfo {author} {\bibfnamefont {A.~P.}\ \bibnamefont {Vieira}}, \
  and\ \bibinfo {author} {\bibfnamefont {S.~R.}\ \bibnamefont {Salinas}},\
  }\href {https://doi.org/10.1103/PhysRevE.83.011701} {\bibfield  {journal}
  {\bibinfo  {journal} {Physical Review E}\ }\textbf {\bibinfo {volume} {83}},\
  \bibinfo {pages} {011701} (\bibinfo {year} {2011})}\BibitemShut {NoStop}%
\bibitem [{\citenamefont {Liarte}\ and\ \citenamefont
  {Salinas}(2012)}]{liarte2012enhancement}%
  \BibitemOpen
  \bibfield  {author} {\bibinfo {author} {\bibfnamefont {D.~B.}\ \bibnamefont
  {Liarte}}\ and\ \bibinfo {author} {\bibfnamefont {S.~R.}\ \bibnamefont
  {Salinas}},\ }\href {https://doi.org/10.1007/s13538-012-0085-y} {\bibfield
  {journal} {\bibinfo  {journal} {Brazilian Journal of Physics}\ }\textbf
  {\bibinfo {volume} {42}},\ \bibinfo {pages} {261} (\bibinfo {year}
  {2012})}\BibitemShut {NoStop}%
\bibitem [{\citenamefont {Sauerwein}\ and\ \citenamefont {{de
  Oliveira}}(2016)}]{Sauerwein2016}%
  \BibitemOpen
  \bibfield  {author} {\bibinfo {author} {\bibfnamefont {R.~A.}\ \bibnamefont
  {Sauerwein}}\ and\ \bibinfo {author} {\bibfnamefont {M.~J.}\ \bibnamefont
  {{de Oliveira}}},\ }\href {\doibase 10.1063/1.4948627} {\bibfield  {journal}
  {\bibinfo  {journal} {The Journal of Chemical Physics}\ }\textbf {\bibinfo
  {volume} {144}},\ \bibinfo {pages} {194904} (\bibinfo {year}
  {2016})}\BibitemShut {NoStop}%
\bibitem [{\citenamefont {Nascimento}\ \emph {et~al.}(2016)\citenamefont
  {Nascimento}, \citenamefont {Vieira},\ and\ \citenamefont
  {Salinas}}]{nascimento2016lattice}%
  \BibitemOpen
  \bibfield  {author} {\bibinfo {author} {\bibfnamefont {E.~S.}\ \bibnamefont
  {Nascimento}}, \bibinfo {author} {\bibfnamefont {A.~P.}\ \bibnamefont
  {Vieira}}, \ and\ \bibinfo {author} {\bibfnamefont {S.~R.}\ \bibnamefont
  {Salinas}},\ }\href {https://doi.org/10.1007/s13538-016-0451-2} {\bibfield
  {journal} {\bibinfo  {journal} {Brazilian Journal of Physics}\ }\textbf
  {\bibinfo {volume} {46}},\ \bibinfo {pages} {664} (\bibinfo {year}
  {2016})}\BibitemShut {NoStop}%
\bibitem [{\citenamefont {Petri}\ and\ \citenamefont
  {Salinas}(2018)}]{petri2018field}%
  \BibitemOpen
  \bibfield  {author} {\bibinfo {author} {\bibfnamefont {A.}~\bibnamefont
  {Petri}}\ and\ \bibinfo {author} {\bibfnamefont {S.~R.}\ \bibnamefont
  {Salinas}},\ }\href {https://doi.org/10.1080/02678292.2017.1404151}
  {\bibfield  {journal} {\bibinfo  {journal} {Liquid Crystals}\ }\textbf
  {\bibinfo {volume} {45}},\ \bibinfo {pages} {980} (\bibinfo {year}
  {2018})}\BibitemShut {NoStop}%
\bibitem [{\citenamefont {Rodrigues}\ \emph {et~al.}(2020)\citenamefont
  {Rodrigues}, \citenamefont {Vieira},\ and\ \citenamefont
  {Salinas}}]{rodrigues2020magnetic}%
  \BibitemOpen
  \bibfield  {author} {\bibinfo {author} {\bibfnamefont {D.~D.}\ \bibnamefont
  {Rodrigues}}, \bibinfo {author} {\bibfnamefont {A.~P.}\ \bibnamefont
  {Vieira}}, \ and\ \bibinfo {author} {\bibfnamefont {S.~R.}\ \bibnamefont
  {Salinas}},\ }\href {https://doi.org/10.3390/cryst10080632} {\bibfield
  {journal} {\bibinfo  {journal} {Crystals}\ }\textbf {\bibinfo {volume}
  {10}},\ \bibinfo {pages} {632} (\bibinfo {year} {2020})}\BibitemShut
  {NoStop}%
\bibitem [{\citenamefont {{dos Santos}}\ \emph {et~al.}(2021)\citenamefont
  {{dos Santos}}, \citenamefont {Vieira}, \citenamefont {Salinas},\ and\
  \citenamefont {Andrade}}]{dosSantos2021}%
  \BibitemOpen
  \bibfield  {author} {\bibinfo {author} {\bibfnamefont {C.~T.~G.}\
  \bibnamefont {{dos Santos}}}, \bibinfo {author} {\bibfnamefont {A.~P.}\
  \bibnamefont {Vieira}}, \bibinfo {author} {\bibfnamefont {S.~R.}\
  \bibnamefont {Salinas}}, \ and\ \bibinfo {author} {\bibfnamefont {R.~F.~S.}\
  \bibnamefont {Andrade}},\ }\href {\doibase 10.1103/PhysRevE.103.032111}
  {\bibfield  {journal} {\bibinfo  {journal} {Phys. Rev. E}\ }\textbf {\bibinfo
  {volume} {103}},\ \bibinfo {pages} {032111} (\bibinfo {year}
  {2021})}\BibitemShut {NoStop}%
\bibitem [{\citenamefont {Salinas}\ and\ \citenamefont
  {Nascimento}(2017)}]{SalinasNascimento2017}%
  \BibitemOpen
  \bibfield  {author} {\bibinfo {author} {\bibfnamefont {S.~R.}\ \bibnamefont
  {Salinas}}\ and\ \bibinfo {author} {\bibfnamefont {E.~S.}\ \bibnamefont
  {Nascimento}},\ }\href {\doibase 10.1080/15421406.2017.1402640} {\bibfield
  {journal} {\bibinfo  {journal} {Molecular Crystals and Liquid Crystals}\
  }\textbf {\bibinfo {volume} {657}},\ \bibinfo {pages} {27} (\bibinfo {year}
  {2017})},\ \Eprint
  {http://arxiv.org/abs/https://doi.org/10.1080/15421406.2017.1402640}
  {https://doi.org/10.1080/15421406.2017.1402640} \BibitemShut {NoStop}%
\bibitem [{\citenamefont {Uzunov}(1993)}]{Uzunov_Book}%
  \BibitemOpen
  \bibfield  {author} {\bibinfo {author} {\bibfnamefont {D.~I.}\ \bibnamefont
  {Uzunov}},\ }\href {\doibase 10.1142/1214} {\emph {\bibinfo {title}
  {Introduction to the Theory of Critical Phenomena}}}\ (\bibinfo  {publisher}
  {WORLD SCIENTIFIC},\ \bibinfo {year} {1993})\ \Eprint
  {http://arxiv.org/abs/https://www.worldscientific.com/doi/pdf/10.1142/1214}
  {https://www.worldscientific.com/doi/pdf/10.1142/1214} \BibitemShut {NoStop}%
\bibitem [{\citenamefont {{de Oliveira}}(2013)}]{Mario_Book}%
  \BibitemOpen
  \bibfield  {author} {\bibinfo {author} {\bibfnamefont {M.~J.}\ \bibnamefont
  {{de Oliveira}}},\ }\href@noop {} {\emph {\bibinfo {title} {Equilibrium
  Thermodynamics}}},\ \bibinfo {edition} {1st}\ ed.\ (\bibinfo  {publisher}
  {Springer-Verlag Berlin Heidelberg},\ \bibinfo {year} {2013})\BibitemShut
  {NoStop}%
\bibitem [{\citenamefont {Luders}\ \emph {et~al.}(2021)\citenamefont {Luders},
  \citenamefont {Arcolezi}, \citenamefont {Pereira}, \citenamefont {Braga},
  \citenamefont {Santos}, \citenamefont {Simões}, \citenamefont {Kimura},
  \citenamefont {Sampaio},\ and\ \citenamefont {Palangana}}]{Luders2021}%
  \BibitemOpen
  \bibfield  {author} {\bibinfo {author} {\bibfnamefont {D.}~\bibnamefont
  {Luders}}, \bibinfo {author} {\bibfnamefont {G.}~\bibnamefont {Arcolezi}},
  \bibinfo {author} {\bibfnamefont {M.}~\bibnamefont {Pereira}}, \bibinfo
  {author} {\bibfnamefont {W.}~\bibnamefont {Braga}}, \bibinfo {author}
  {\bibfnamefont {O.}~\bibnamefont {Santos}}, \bibinfo {author} {\bibfnamefont
  {M.}~\bibnamefont {Simões}}, \bibinfo {author} {\bibfnamefont
  {N.}~\bibnamefont {Kimura}}, \bibinfo {author} {\bibfnamefont
  {A.}~\bibnamefont {Sampaio}}, \ and\ \bibinfo {author} {\bibfnamefont
  {A.}~\bibnamefont {Palangana}},\ }\href {\doibase
  10.1080/02678292.2020.1836279} {\bibfield  {journal} {\bibinfo  {journal}
  {Liquid Crystals}\ }\textbf {\bibinfo {volume} {48}},\ \bibinfo {pages} {974}
  (\bibinfo {year} {2021})}\BibitemShut {NoStop}%
\bibitem [{\citenamefont {Skutnik}\ \emph {et~al.}(2020)\citenamefont
  {Skutnik}, \citenamefont {Geier},\ and\ \citenamefont
  {Schoen}}]{Skutnik2020}%
  \BibitemOpen
  \bibfield  {author} {\bibinfo {author} {\bibfnamefont {R.~A.}\ \bibnamefont
  {Skutnik}}, \bibinfo {author} {\bibfnamefont {I.~S.}\ \bibnamefont {Geier}},
  \ and\ \bibinfo {author} {\bibfnamefont {M.}~\bibnamefont {Schoen}},\ }\href
  {\doibase 10.1080/00268976.2020.1726520} {\bibfield  {journal} {\bibinfo
  {journal} {Molecular Physics}\ }\textbf {\bibinfo {volume} {118}},\ \bibinfo
  {pages} {e1726520} (\bibinfo {year} {2020})}\BibitemShut {NoStop}%
\bibitem [{\citenamefont {Amaral}\ \emph {et~al.}(1997)\citenamefont {Amaral},
  \citenamefont {Filho}, \citenamefont {Taddei},\ and\ \citenamefont
  {Vila-Romeu}}]{Amaral1997}%
  \BibitemOpen
  \bibfield  {author} {\bibinfo {author} {\bibfnamefont {L.~Q.}\ \bibnamefont
  {Amaral}}, \bibinfo {author} {\bibfnamefont {O.~S.}\ \bibnamefont {Filho}},
  \bibinfo {author} {\bibfnamefont {G.}~\bibnamefont {Taddei}}, \ and\ \bibinfo
  {author} {\bibfnamefont {N.}~\bibnamefont {Vila-Romeu}},\ }\href {\doibase
  10.1021/la9700073} {\bibfield  {journal} {\bibinfo  {journal} {Langmuir}\
  }\textbf {\bibinfo {volume} {13}},\ \bibinfo {pages} {5016} (\bibinfo {year}
  {1997})}\BibitemShut {NoStop}%
\bibitem [{\citenamefont {Arfken}\ and\ \citenamefont
  {Weber}(2005)}]{arfken2005}%
  \BibitemOpen
  \bibfield  {author} {\bibinfo {author} {\bibfnamefont {G.}~\bibnamefont
  {Arfken}}\ and\ \bibinfo {author} {\bibfnamefont {H.~J.}\ \bibnamefont
  {Weber}},\ }\href@noop {} {\emph {\bibinfo {title} {Mathematical Methods for
  Physicists, 6th ed.}}}\ (\bibinfo  {publisher} {Elsevier Academic Press},\
  \bibinfo {year} {2005})\BibitemShut {NoStop}%
\end{thebibliography}%

\end{document}